\documentclass[10pt,journal,letterpaper,compsoc,twoside]{IEEEtran}

\ifCLASSOPTIONcompsoc
  \usepackage[nocompress]{cite}
\else
  \usepackage{cite}
\fi

\ifCLASSINFOpdf
\else
\fi
\ifCLASSOPTIONcompsoc
  \usepackage[caption=false,font=footnotesize,labelfont=sf,textfont=sf]{subfig}
\else
  \usepackage[caption=false,font=footnotesize]{subfig}
\fi

\makeatletter
\newif\if@restonecol
\makeatother

\usepackage{graphicx}
\usepackage{algorithm}
\usepackage{algorithmicx}
\usepackage[noend]{algpseudocode}

\usepackage{comment}
\usepackage{epstopdf}
\usepackage{epsfig}
\usepackage{tabularx}
\usepackage{enumitem}
\usepackage{amssymb,amsmath,amsthm}
\usepackage{url}

\usepackage{textcomp}

\newcommand{\ie}{{\em i.e.}}
\newcommand{\eg}{{\em e.g.}}
\newcommand{\et}{{\em et al.}}

\begin{document}
\title{Achieving Data Truthfulness and Privacy Preservation in Data Markets}

\author{Chaoyue~Niu,~\IEEEmembership{Student~Member,~IEEE,}
        Zhenzhe~Zheng,~\IEEEmembership{Student~Member,~IEEE,}
        Fan~Wu,~\IEEEmembership{Member,~IEEE,}
        Xiaofeng~Gao,~\IEEEmembership{Member,~IEEE,}
        and~Guihai~Chen,~\IEEEmembership{Senior~Member,~IEEE}
\IEEEcompsocitemizethanks{\IEEEcompsocthanksitem The authors are with the Shanghai Key Laboratory of Scalable Computing and Systems, Department of Computer Science and Engineering, Shanghai~Jiao~Tong~University, Shanghai, 200000, China.\protect\\
E-mail: \{rvincency, zhengzhenzhe220\}@gmail.com; \{fwu, gao-xf, gchen\}@cs.sjtu.edu.cn}
\thanks{Manuscript received 11 Aug. 2017; revised 3 Dec. 2017; accepted 19 Mar. 2018. Date of publication XXXX 2018; date of current version 13 May 2018.\protect\\
Recommended for acceptance by J. Chen.\protect\\
For information on obtaining reprints of this article, please send e-mail to reprints@ieee.org, and reference the Digital Object Identifier below.\protect\\
Digital Object Identifier no. 10.1109/TKDE.2018.2822727
}
}


%
%

\markboth{IEEE Transactions on Knowledge and Data Engineering,~Vol.~XX, No.~XX, XXXX~2018}%
{Niu \MakeLowercase{\textit{et al.}}: Achieving Data Truthfulness and Privacy Preservation in Data Markets}
%



\IEEEtitleabstractindextext{%
\begin{abstract}
As a significant business paradigm, many online information platforms have emerged to satisfy society's needs for person-specific data, where a service provider collects raw data from data contributors, and then offers value-added data services to data consumers. However, in the data trading layer, the data consumers face a pressing problem, \ie, how to verify whether the service provider has truthfully collected and processed data? Furthermore, the data contributors are usually unwilling to reveal their sensitive personal data and real identities to the data consumers. In this paper, we propose TPDM, which efficiently integrates data \underline{T}ruthfulness and \underline{P}rivacy preservation in \underline{D}ata \underline{M}arkets. TPDM is structured internally in an Encrypt-then-Sign fashion, using partially homomorphic encryption and identity-based signature. It simultaneously facilitates batch verification, data processing, and outcome verification, while maintaining identity preservation and data confidentiality. We also instantiate TPDM with a profile matching service and a distribution fitting service, and extensively evaluate their performances on Yahoo! Music ratings dataset and 2009 RECS dataset, respectively. Our analysis and evaluation results reveal that TPDM achieves several desirable properties, while incurring low computation and communication overheads when supporting large-scale data markets.
\end{abstract}

}

\maketitle

\IEEEdisplaynontitleabstractindextext

%
\IEEEpeerreviewmaketitle


\IEEEraisesectionheading{\section{Introduction}\label{sec:introduction}}


\IEEEPARstart{I}{n} the era of big data, society has developed an insatiable appetite for sharing personal data. Realizing the potential of personal data's economic value in decision making and user experience enhancement, several open information platforms have emerged to enable person-specific data to be exchanged on the Internet~\cite{GNIP,datasift,link:datacoup,link:citizenme,tech:14:ftc}. For example, Gnip, which is Twitter's enterprise API platform, collects social media data from Twitter users, mines deep insights into customized audiences, and provides data analysis solutions to more than 95$\%$ of the Fortune 500~\cite{GNIP}.


However, there exists a critical security problem in these market-based platforms, \ie, it is difficult to guarantee the truthfulness in terms of data collection and data processing, especially when the privacies of the data contributors are needed to be preserved. Let's examine the role of a pollster in the presidential election as follows. As a reliable source of intelligence, the Gallup Poll~\cite{Gallup_Poll} uses impeccable data to assist presidential candidates in identifying and monitoring economic and behavioral indicators. In this scenario, simultaneously ensuring data truthfulness and preserving privacy require the Gallup Poll to convince the presidential candidates that those indicators are derived from live interviews without leaking any interviewer's real identity (\eg, social security number) or the content of her interview. If raw data sets for drawing these indicators are mixed with even a small number of bogus or synthetic samples, it will exert bad influence on the final election result.





Ensuring data truthfulness and protecting the privacies of data contributors are both important to the long term healthy development of data markets. On one hand, the ultimate goal of the service provider in a data market is to maximize her profit. Therefore, in order to minimize the expenditure for data acquisition, an opportunistic way for the service provider is to mingle some bogus or synthetic data into the raw data set. Yet, to reduce operation cost, a cunning service provider may provide data services based on a subset of the whole raw data set, or even return a fake result without processing the data from designated sources. However, if such speculative and illegal behaviors cannot be identified and prohibited, it will cause heavy losses to data consumers, and thus destabilize the data market. On the other hand, while unleashing the power of personal data, it is the bottom line of every business to respect the privacies of data contributors. The debacle, which follows AOL's public release of ``anonymized'' search records of its customers, highlights the potential risk to individuals in sharing personal data with private companies~\cite{barbaro2006face}. Besides, according to the survey report of 2016 TRUSTe/NCSA Consumer Privacy Infographic - US Edition~\cite{link:privacy:2016survey}, $89\%$ of consumers say they avoid companies that do not respect privacy. Therefore, the content of raw data should not be disclosed to the data consumers to guarantee data confidentiality, even if the real identities of the data contributors are hidden.

To integrate data truthfulness and privacy preservation in a practical data market, there are four major challenges. The first and the thorniest design challenge is that verifying the truthfulness of data collection and preserving the privacy seem to be contradictory objectives. Ensuring the truthfulness of data collection allows the data consumers to verify the validities of data contributors' identities and the content of raw data, whereas privacy preservation tends to prevent them from learning these confidential contents. Specifically, the property of non-repudiation in classical digital signature schemes implies that the signature is unforgeable, and any third party is able to verify the authenticity of a data submitter using her public key and the corresponding digital certificate, \ie, the truthfulness of data collection in our model. However, the verification in digital signature schemes requires the knowledge of raw data, and can easily leak a data contributor's real identity~\cite{RenK2006}. Regarding a message authentication code (MAC), each pair of a data contributor and a data consumer need to agree on a shared secret key, which is unpractical in data markets.


Yet, another challenge comes from data processing, which makes verifying the truthfulness of data collection even harder. Nowadays, more and more data markets provide data services rather than directly offering raw data. The following reasons account for this trend: 1) For the data contributors, they have severe privacy concerns~\cite{link:privacy:2016survey}. Nevertheless, the service-based trading mode, which has hidden the sensitive raw data, alleviates their concerns; 2) For the service provider, comprehensive and insightful data services can bring in more profits~\cite{DataMarket2011}; 3) For the data consumers, data copyright infringement~\cite{datalawyer:15sigmod} is serious. However, such a data trading mode differs from most of conventional data sharing scenarios, \eg, data publishing~\cite{Kalnis:2011:PPDP}. Besides, the data services, as the results of data processing, may no longer be semantically consistent with the raw data~\cite{Fung2010PDP}, which makes the data consumers hard to believe the truthfulness of data collection. In addition, the digital signatures on raw data become invalid for the data services, which discourages the data consumers from doing verification as mentioned above. Moreover, although data provenance~\cite{ICDE2013:Provenance} helps to determine the derivation histories of data processing results, it cannot guarantee the truthfulness of data collection. While knowledge provenance~\cite{KnowledgeProvenance:2011}, an enhanced version of data provenance, tackles the deficiency of data provenance, but it breaks the property of identity preservation.

The third challenge lies in how to guarantee the truthfulness of data processing, under the information asymmetry between the data consumers and the service provider due to data confidentiality. In particular, to ensure data confidentiality against the data consumers, the service provider can employ a conventional symmetric/asymmetric cryptosystem, and let the data contributors encrypt their raw data. Unfortunately, a hidden problem arisen is that the data consumers fail to verify the correctness and completeness of returned data services. Even worse, some greedy service providers may exploit this vulnerability to reduce operation cost during the execution of data processing, \eg, they might return an incomplete data service without processing the whole data set, or even return an outright fake result without processing the data from designated data sources.

Last but not least, the fourth design challenge is the efficiency requirement of data markets, especially for data acquisition, \ie, the service provider should be able to collect data from a large number of data contributors with low latency. Due to the timeliness of some kinds of person-specific data, the service provider has to periodically collect fresh raw data to meet the diverse demands of high-quality data services. For example, 25 billion data collection activities take place on Gnip every day~\cite{GNIP}. Meanwhile, the service provider needs to verify data authentication and data integrity. One basic approach is to let each data contributor sign her raw data. However, classical digital signature schemes, which verify the received signatures one after another, may fail to satisfy the stringent time requirement of data markets. Furthermore, the maintenance of digital certificates under the traditional Public Key Infrastructure (PKI) also incurs significant communication overhead. Under such circumstances, verifying a large number of signatures sequentially certainly becomes the processing bottleneck at the service provider.

In this paper, by jointly considering above four challenges, we propose TPDM, which achieves both data \underline{T}ruthfulness and \underline{P}rivacy preservation in \underline{D}ata \underline{M}arkets. TPDM first exploits partially homomorphic encryption to construct a ciphertext space, which enables the service provider to launch data services and the data consumers to verify the truthfulness of data processing, while maintaining data confidentiality. In contrast to classical digital signature schemes, which are operated over plaintexts, our new identity-based signature scheme is conducted in the ciphertext space. Furthermore, each data contributor's signature is derived from her real identity, and is unforgeable against the service provider or other external attackers. This appealing property can convince the data consumers that the service provider has truthfully collected data. To reduce the latency caused by verifying a bulk of signatures, we propose a two-layer batch verification scheme, which is built on the bilinear property of admissible pairing. At last, TPDM realizes identity preservation and revocability by carefully adopting ElGamal encryption and introducing a semi-honest registration center.

We summarize our key contributions as follows.

$\bullet$ To the best of our knowledge, TPDM is the first secure mechanism for data markets achieving both data truthfulness and privacy preservation.


$\bullet$ TPDM is structured internally in a way of Encrypt-then-Sign using partially homomorphic encryption and identity-based signature. It enforces the service provider to truthfully collect and process real data. Besides, TPDM incorporates a two-layer batch verification scheme with an efficient outcome verification scheme, which can drastically reduce computation overhead.


$\bullet$ We instructively instantiate TPDM with two kinds of practical data services, namely profile matching and distribution fitting. Besides, we implement these two concrete data markets, and extensively evaluate their performances on Yahoo! Music ratings dataset and 2009 RECS dataset. Our analysis and evaluation results reveal that TPDM achieves good effectiveness and efficiency in large-scale data markets. Specifically, for profile matching, when supporting as many as 1 million data contributors in one session of data acquisition, the computation and communication overheads at the service provider are 0.930s and 0.235KB per matching with 10 attributes in each profile, respectively. In addition, the outcome verification's overhead per matching is only $1.17\%$ of the original similarity evaluation's cost.




The remainder of this paper is organized as follows. In Section~\ref{system_attack_models}, we introduce system model, adversary model, and technical preliminary. We show the detailed design of TPDM in Section~\ref{TPDM}, and analyze its security in Section~\ref{securityanalysis}. In Section~\ref{sec:two:datamarkets}, we elaborate on the applications of TPDM to profile matching and distribution fitting. The evaluation results are presented in Section~\ref{evaluation}. We briefly review related work in Section~\ref{related_work}. We conclude the paper in Section~\ref{conclusion}.

\section{Preliminaries}\label{system_attack_models}
In this section, we first describe a general system model for data markets. We then introduce the adversary model, and present corresponding security requirements on the design. We finally review technical preliminary.


\subsection{System Model}

\begin{figure}[!t]
\centering
\includegraphics [width = 0.8\columnwidth]{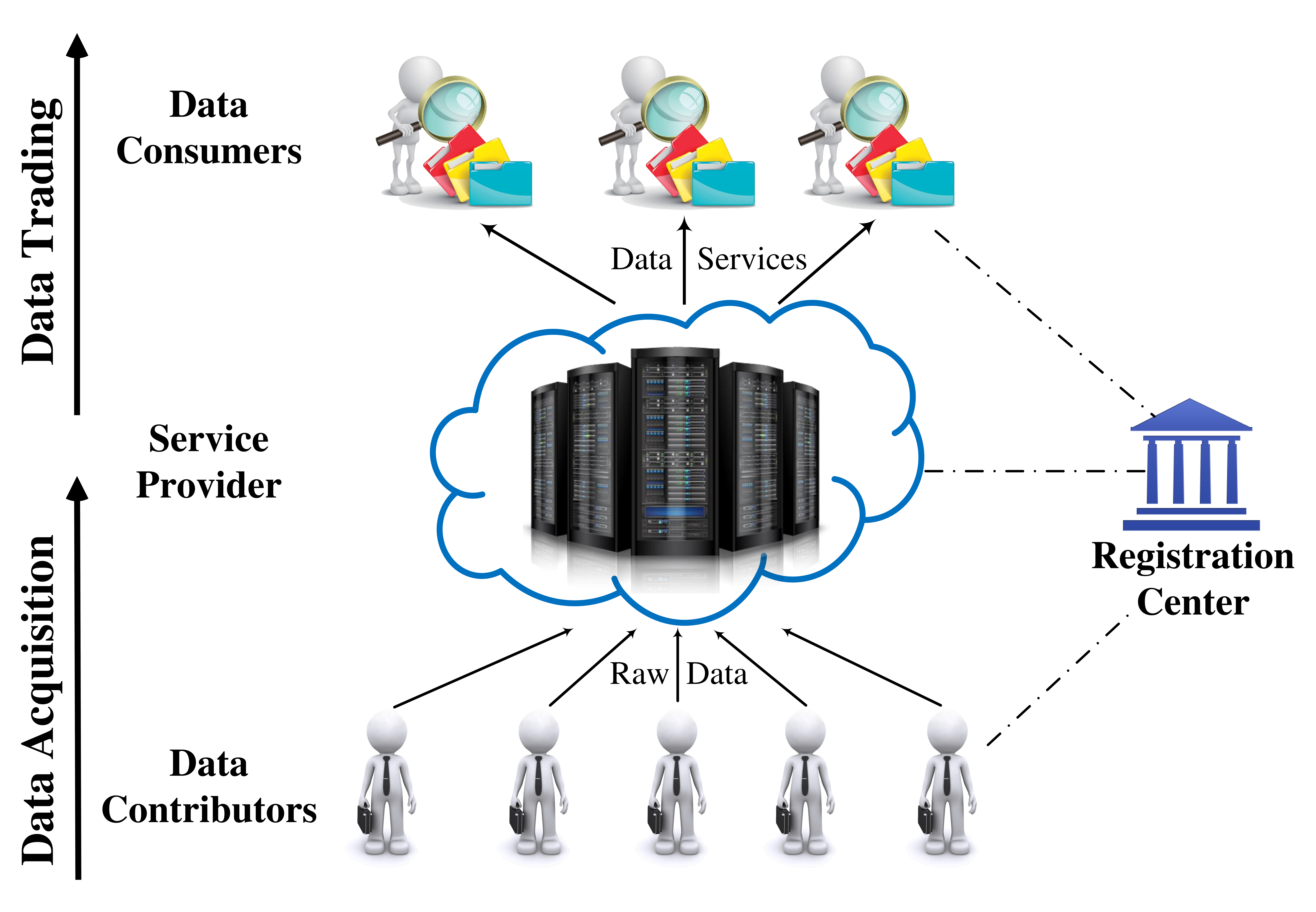} 
\caption{A two-layer system model for data markets.}
\label{fig:systemmodel}
\end{figure}

As shown in Fig.~\ref{fig:systemmodel}, we consider a two-layer system model for data markets. The model has a data acquisition layer and a data trading layer. There are four major kinds of entities, including data contributors, a service provider, data consumers, and a registration center.

In the data acquisition layer, the service provider procures massive raw data from the data contributors, such as social network users, mobile smart devices, smart meters, and so on. In order to incentivize more data contributors to actively submit high-quality data, the service provider needs to reward those valid ones to compensate their data collection costs. For the sake of security, each registered data contributor is equipped with a tamper-proof device. The tamper-proof device can be implemented in the form of either specific hardware~\cite{JCS07:hardware:tamperproofdevice} or software~\cite{SPECS:2011:VANET}. It prevents any adversary from extracting the information stored in the device, including cryptographic keys, codes, and data.


We consider that the service provider is cloud based, and has abundant computing resources, network bandwidths, and storage space. Besides, she tends to offer semantically rich and value-added data services to data consumers rather than directly revealing sensitive raw data, \eg, social network analyses, probability distributions, personalized recommendations, and aggregate statistics.





The registration center maintains an online database of registrations, and assigns each registered data contributor an identity and a password to activate the tamper-proof device. Besides, she maintains an official website, called certificated bulletin board~\cite{jour:jnca17:era,proc:eurocrypt97:cramer:cbb}, on which the legitimate system participants can publish essential information, \eg, whitelists, blacklists, resubmit-lists, and reward-lists of data contributors. Yet, another duty of the registration center is to set up the parameters for a signature scheme and a cryptosystem. To avoid being a single point of failure or bottleneck, redundant registration centers, which have identical functionalities and databases, can be installed.

\subsection{Adversary Model}\label{adversary:security}
\begin{figure*}[t]
\centering
\includegraphics[width = 1.66\columnwidth]{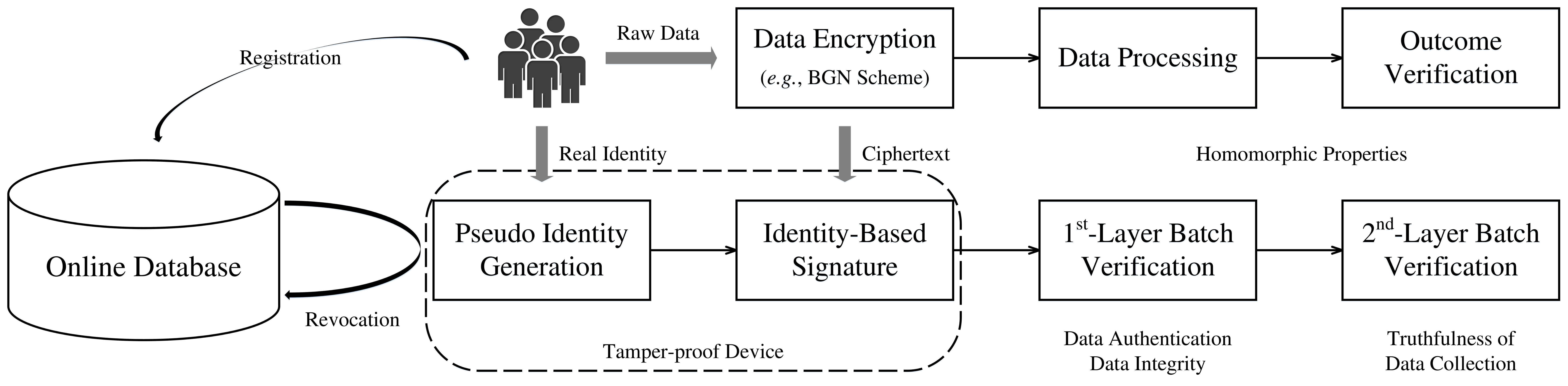}
\caption{System architecture of TPDM.}\label{fig:designrationale}
\end{figure*}


In this section, we focus on attacks in practical data markets, and define corresponding security requirements.

First, we consider that a malicious data contributor or an external attacker may impersonate other legitimate data contributors to submit possibly bogus raw data. Besides, some malicious attackers may deliberately modify raw data during submission. Hence, the service provider needs to confirm that raw data are indeed sent unaltered by registered data contributors, \ie, to guarantee \emph{data authentication and data integrity} in the data acquisition layer.


Second, the service provider in the data market might be greedy, and attempts to maximize her profit by launching the following two types of attacks:
\begin{itemize}
\item\emph{Partial data collection:} To cut down the expenditure on data acquisition, the service provider may insert bogus data into the raw data set.
\item\emph{No/Partial data processing:} To reduce the operation cost, the service provider may try to return a fake result without processing the data from designated sources, or to provide data services based on a subset of the whole raw data set.
\end{itemize}
On one hand, to counter partial data collection attack, each data consumer should be enabled to verify whether raw data are really provided by registered data contributors, \ie, \emph{truthfulness of data collection} in the data trading layer. On the other hand, the data consumer should have the capability to verify the \emph{correctness} and \emph{completeness} of a returned data service in order to combat no/partial data processing attack. We here use the term \emph{truthfulness of data processing} in the data trading layer to represent the integrated requirement of correctness and completeness of data processing results.

Third, we assume that some honest-but-curious data contributors, the service provider, the data consumers, and external attackers, \eg, eavesdroppers, may glean sensitive information from raw data, and recognize real identities of data contributors for illegal purposes, \eg, an attacker can infer a data contributor's home location from her GPS records. 
Hence, raw data of a data contributor should be kept secret from these system participants, \ie, \emph{data confidentiality}. Besides, an outside observer cannot reveal a data contributor's real identity by analysing data sets sent by her, \ie, \emph{identity preservation}. 

Fourth, a minority of data contributors may try to behave illegally, \eg, launching attacks as mentioned above, if there is no punishment. To prevent this threat, the registration center should have the ability to retrieve a data contributor's real identity, and revoke it from further usage, when her signature is in dispute, \ie, \emph{traceability and revocability}.

Last but not least, the semi-honest registration center may misbehave by trying to link a data contributor's real identity with her raw data. Besides, if there is no detection or verification in the cryptosystem, she may deliberately corrupt the decrypted results. However, to guarantee full side information protection, the requirement on the registration center is that she cannot leak decrypted samples to irrelevant system participants. Moreover, she is required to perform an acknowledged number of decryptions in a specific data service~\cite{HE:Paillier:ccs2011}, which should be publicly posted on the certificated bulletin board.

\subsection{Admissible Pairing}\label{preliminaries}
In this section, we introduce admissible pairing, which is the basis of our design.

The multiplicative cyclic groups $\mathbb{G}_1, \mathbb{G}_2$, and $\mathbb{G}_T$ are of the same prime order $q$. Let $g_1$ be a generator of $\mathbb{G}_1$, and $g_2$ be a generator of $\mathbb{G}_2$. An asymmetric bilinear map is a map $\hat{e}: \mathbb{G}_1\times\mathbb{G}_2\rightarrow\mathbb{G}_T$ with the following three properties:
\begin{itemize}
  \item \emph{Bilinearity}: $\forall X, Y\in\mathbb{G}_1, \forall Z \in\mathbb{G}_2, \forall a, b\in\mathbb{Z}{_q^*}$,
$$\hat{e}(X^a,Z^b)=\hat{e}(X,Z)^{ab}.$$
In addition,
$$\hat{e}(XY, Z)= \hat{e}(X, Z) \cdot \hat{e}(Y, Z).$$
  \item \emph{Non-degeneracy}: $\hat{e}(g_1,g_2) \neq 1_{\mathbb{G}_T}.$ 
  \item \emph{Computability}: Given $X \in\mathbb{G}_1, Z \in\mathbb{G}_2$, there exists an efficient algorithm to compute $\hat{e}(X,Z)$.
\end{itemize}


We call such a bilinear map $\hat{e}$ an admissible pairing, which can be constructed based on elliptic curves with modified Weil~\cite{IBE2001} or Tate pairing~\cite{TatePairing}. Each operation for computing $\hat{e}(X,Z)$ is called pairing operation. The group that possesses such a map $\hat{e}$ is called a bilinear group, where the Decisional Diffie-Hellman (DDH) problem is easy, while the Computational Diffie-Hellman (CDH) problem is hard~\cite{IBE2001}. For example, given $(g_1, {g_1}^a, {g_1}^b)$ for unknown $(a, b)$, it is computationally intractable to compute ${g_1}^{ab}$.

\section{Design of TPDM}\label{TPDM}
In this section, we propose TPDM, which integrates data truthfulness and privacy preservation in data markets.

\subsection{Design Rationales}\label{Designoverview}

Using the terminology from the signcryption scheme~\cite{AN2002SE}, TPDM is structured internally in a way of Encrypt-then-Sign, using partially homomorphic encryption and identity-based signature. It enforces the service provider to truthfully collect and process real data. The essence of TPDM is to first synchronize data processing and signature verification into the same ciphertext space, and then to tightly integrate data processing with outcome verification via the homomorphic properties. With the help of the architectural overview in Fig.~\ref{fig:designrationale}, we illustrate the design rationales as follows.


\textbf{Space Construction.} The thorniest problem is how to enable the data consumer to verify the validnesses of signatures, while maintaining data confidentiality. If the signature scheme is applied to the plaintext space, the data consumer needs to know the content of raw data for verification. However, if we employ a conventional public key encryption scheme to construct the ciphertext space, the service provider has to decrypt and then process the data. Even worse, such a construction is vulnerable to the no/partial data processing attack, because the data consumer, only knowing the ciphertexts, fails to verify the correctness and completeness of the data service. Thus, the greedy service provider may reduce operation cost, by returning a fake result or manipulating the inputs of data processing. Therefore, we turn to the partially homomorphic cryptosystems for encryption, whose properties facilitate both data processing and outcome verification on the ciphertexts.

\textbf{Batch Verification.} After constructing the ciphertext space, we can let each data contributor digitally sign her encrypted raw data. Given the ciphertext and the signature, the service provider is able to verify data authentication and data integrity. Besides, we can treat the data consumer as a third party to verify the truthfulness of data collection. However, an immediate question arisen is that the sequential verification schema may fail to meet the stringent time requirement of large-scale data markets. In addition, the maintenance of digital certificates also incurs significant communication overhead. To tackle these two problems, we propose an identity-based signature scheme, which supports two-layer batch verifications, while incurring small computation and communication overheads.
\textbf{Breach Detection.} Yet, another problem in existing identity-based signature schemes is that the real identities are viewed as public parameters, and are not well protected. On the other hand, if all the real identities are hidden, none of the misbehaved data contributors can be identified. To meet these two seemly contradictory requirements, we employ ElGamal encryption to generate pseudo identities for registered data contributors, and introduce a new third party, called registration center. Specifically, the registration center, who owns the private key, is the only authorized party to retrieve the real identities, and to revoke those malicious accounts from further usage.

\subsection{Design Details}\label{DesignDetails}
Following the guidelines given above, we now introduce TPDM in detail. TPDM consists of 5 phases: initialization, signing key generation, data submission, data processing and verifications, and tracing and revocation.


\vspace{0.4em}
\noindent \textbf{Phase \uppercase\expandafter{\romannumeral1}: Initialization}
\vspace{0.4em}

We assume that the registration center sets up the system parameters at the beginning of data trading as follows:

$\bullet$~The registration center chooses three multiplicative cyclic groups $\mathbb{G}_1$, $\mathbb{G}_2$, and $\mathbb{G}_T$ with the same prime order $q$. Besides, $g_1$ is a generator of $\mathbb{G}_1$, and $g_2$ is a generator of $\mathbb{G}_2$. Moreover, these three cyclic groups compose an admissible pairing $\hat{e}: \mathbb{G}_1 \times \mathbb{G}_2 \rightarrow \mathbb{G}_T$.

$\bullet$~The registration center randomly picks $s_1, s_2 \in \mathbb{Z}{_q^*}$ as her two master keys, and then computes
\begin{align}
P_0 = {g_1}^{s_1}, P_1 = {g_2}^{s_1},\ \text{and}\ P_2 = {g_2}^{s_2}
\end{align}
as public keys. The master keys $s_1, s_2$ are preloaded into each registered data contributor's tamper-proof device.

$\bullet$~The registration center sets up parameters for a partially homomorphic cryptosystem: a private key $\mathcal{SK}$, a public key $\mathcal{PK}$, an encryption scheme $E(\cdot)$, and a decryption scheme $D(\cdot)$.


$\bullet$~To activate the tamper-proof device, each registered data contributor $o_i$ is assigned with a ``real'' identity $\emph{RID}_i\in\mathbb{G}_1$ and a password $\emph{PW}_i$. Here, $\emph{RID}_i$ uniquely identifies $o_i$, while $\emph{PW}_i$ is required in the access control process.

$\bullet$~The system parameters $$\left\{\hat{e}, \mathbb{G}_1, \mathbb{G}_2, \mathbb{G}_T,\ q,\ g_1,\ g_2,\ P_0,\ P_1,\ P_2,\ \mathcal{PK},\ E(\cdot)\right\}$$ are published on the certificated bulletin board.

\vspace{0.4em}
\noindent \textbf{Phase \uppercase\expandafter{\romannumeral2}: Signing Key Generation}
\vspace{0.4em}

To achieve anonymous authentication in data markets, the tamper-proof device is utilized to generate a pair of pseudo identity $\emph{PID}_i$ and secret key $SK_i$ for each registered data contributor $o_i$:
\begin{align}
\emph{PID}_i &= \langle \emph{PID}_i^1, \emph{PID}_i^2 \rangle = \langle {g_1}^r, \emph{RID}_i\odot {P_0}^r \rangle,\\
SK_i &= \langle SK_i^1, SK_i^2 \rangle = \langle \emph{PID}{_i^1}^{s_1},  H(\emph{PID}{_i^2})^{s_2} \rangle.
\end{align}
Here, $r$ is a per-session random nonce, $\odot$ represents the Exclusive-OR (XOR) operation, and $H$($\cdot$) is a MapToPoint hash function~\cite{IBE2001}, \ie, $H(\cdot): \{0,1\}^*\rightarrow\mathbb{G}_1$. Besides, $\emph{PID}_i$ is an ElGamal encryption~\cite{elgamal1985public} of the real identity $\emph{RID}_i$ over the elliptic curve, while $SK_i$ is generated accordingly by exploiting identity-based encryption (IBE)~\cite{IBE2001}.

\vspace{0.4em}
\noindent \textbf{Phase \uppercase\expandafter{\romannumeral3}: Data Submission}
\vspace{0.4em}

For the submission of raw data, we need to jointly consider several security issues, including confidentiality, authentication, and integrity. To provide data confidentiality, we employ partially homomorphic encryption. Besides, to guarantee data authentication and data integrity, the encrypted raw data should be signed before submission, and also should be verified after reception.

\vspace{0.2em}
\noindent$\blacktriangleright$ \textbf{Data Encryption}
\vspace{0.2em}

Ahead of submission, each data contributor $o_i$ encrypts her raw data $U_i$ to different powers under the public key $\mathcal{PK}$, and gets the ciphertext vector
\begin{equation}
\vec{D}_i = E({U_i}^k)|_{k \in \mathbb{K} \subseteq \mathbb{Z}^+},
\end{equation}
where $\mathbb{K}$ is a set of positive integers, and is determined by the requirements of data services, \eg, the location-based aggregate statistics~\cite{HE:Paillier:ccs2011} may require $\mathbb{K} = \{1\}$, whereas in the fine-grained profile matching~\cite{matching12INFOCOM}, $\mathbb{K} = \{1, 2\}$.

In general, compared with the time-consuming computation on ciphertexts, the evaluation of plaintexts is quite more efficient. Therefore, we let each data contributor encrypt her raw data to different powers, which can benefit an optimization in data processing while incurring a small overhead at each data contributor. 

\vspace{0.2em}
\noindent$\blacktriangleright$ \textbf{Encrypted Data Signing}
\vspace{0.2em}


After encryption, each data contributor $o_i$ computes the signature $\sigma_i$ on the ciphertext vector $\vec{D}_i$ using her secret key:
\begin{equation}\label{eq:signature}
\sigma_i = SK{_i^1} \cdot SK{_i^2}^{h(D_i)},
\end{equation}
where ``$\cdot$'' denotes the group operation in $\mathbb{G}_1$, $h(\cdot)$ is a one-way hash function, \eg, SHA-1~\cite{SHA1}, and $D_i$ is derived by concatenating all the elements of $\vec{D}_i$ together.

Eventually, the data contributor $o_i$ submits her tuple $\langle \emph{PID}_i, \vec{D}_i, \sigma_i\rangle$ to the service provider. Once receiving the tuple, the service provider is required to post the pseudo identity $\emph{PID}_i$ on the certificated bulletin board for fear of receiver-repudiation. In addition, to prevent a registered data contributor from using the same pair of pseudo identity and secret key for multiple times in different sessions of data acquisition (analogous to the replay attack scenario considered in~\cite{SPECS:2011:VANET}), one intuitive way is to let the service provider store those used pseudo identities for duplication check later. Yet, another feasible way is to encapsulate the signing phase into the tamper-proof device.


\vspace{0.4em}
\noindent \textbf{Phase \uppercase\expandafter{\romannumeral4}: Data Processing and Verifications}
\vspace{0.4em}

In this phase, we consider two-layer batch verifications, \ie, verifications conducted by both the service provider and the data consumer. Between the two-layer batch verifications, we introduce data processing and signatures aggregation done by the service provider. At last, we present outcome verification conducted by the data consumer.

Before introducing the verifications, we first discuss the time period $\tau$ of data acquisition. In practice, $\tau$ is determined by the service provider, and is based on the timeliness of different data items. For example, stock data is streaming with a minimum update frequency of 1 minute on Investing~\cite{stockinvesting}, while smart meters collect the electrical usages every 15 minutes~\cite{HE:others:ndss2011}. In what follows, we focus on one time period of data acquisition.

\vspace{0.2em}
\noindent$\blacktriangleright$ \textbf{First-Layer Batch Verification}
\vspace{0.2em}



We assume that the service provider receives a bundle of data tuples from $n$ distinct data contributors, denoted as $\{\langle \emph{PID}_i, \vec{D}{_i}, \sigma_i \rangle|i\in [1, n]\}$, by the end of a time period. To prevent a malicious data contributor from impersonating other legitimate ones to submit possibly bogus data, the service provider needs to verify the validnesses of signatures by checking whether
\begin{align}
\nonumber
&\hat{e}\left(\prod_{i=1}^{n} \sigma_i, g_2\right)\\
=\quad &\hat{e}\left(\prod_{i=1}^{n}\emph{PID}{_i^1}, P_{1}\right) \hat{e}\left(\prod_{i=1}^{n}H(\emph{PID}{_i^2})^{h(D_i)}, P_{2}\right).\label{BV1}
\end{align}

Compared with single signature verification, this batch verification scheme can dramatically reduce the verification latency, especially when verifying a large number of signatures. Since the three pairing operations in Equation~(\ref{BV1}) dominate the overall computation cost, the batch verification time is almost a constant if the time overhead of $n$ MapToPoint hashings and $n$ exponentiations is small enough to be emitted. However, in a practical data market, when the number of data contributors is too large, the expensive pairing operations cannot dominate the verification time. We will expand on this point in Section~\ref{section:computation:overhead}.

\vspace{0.2em}
\noindent$\blacktriangleright$ \textbf{Data Processing and Signatures Aggregation}
\vspace{0.2em}

Instead of directly trading raw data for revenue, more and more service providers tend to trade value-added data services. Typical examples of data services include social network analyses, personalized recommendations, location-based services, and probability distribution fittings.

To facilitate generating a precise and customized strategy in targeted data services, \eg, personalized recommendation and locate-based service, the data consumer also needs to provide her own ciphertext vector $\vec{D}_0$ and a threshold $\delta$. Here, $\vec{D}_0$ is generated from the data consumer's information $V$ as follows:
\begin{eqnarray}\label{eq:d01}
\vec{D}_0 = E(\omega_iV^{\bar{k}_i})|_{\bar{k}_i \in \bar{\mathbb{K}} \subseteq \mathbb{Z}^+, i \in [1, |\bar{\mathbb{K}}|]},
\end{eqnarray}
where $\bar{k}_i, \omega_i$ are parameters determined by a concrete data service. For example, the profile-matching service in Section~\ref{profilematching} requires $\bar{k}_i \in \{1, 2\}$ and $\omega_i \in \{-2, 1\}$.


Now, the service provider can process the collected data as required by the data consumer. We model such a data processing in the plaintext space as
\begin{equation}\label{eq:dataprocessing_p}
\gamma = f\left(V, U_{c_1}, U_{c_2}, \cdots, U_{c_m}\right)
\end{equation}
for generality. Accordingly, $f$ can be equivalently evaluated in the ciphertext space using
\begin{align}
R = E(\gamma) 
  = F(\vec{D}_0, \vec{D}_{c_1}, \vec{D}_{c_2}, \cdots, \vec{D}_{c_m}). \label{eq:dataprocessing_c}
\end{align}
The equivalent transformation from $f$ to $F$ is based on the properties of the partially homomorphic cryptosystem, \eg, homomorphic addition $\oplus$ and homomorphic multiplication $\otimes$, which are arithmetic operations on the ciphertexts that are equivalent to the usual addition and multiplication on the plaintexts, respectively. Hence, only polynomial functions can be computed in a straightforward way. Nevertheless, most non-polynomial functions, \eg, sigmoid and rectified linear activation functions in machine learning, can be well approximated/handled by polynomials~\cite{proc:icml16:cryptonet}. Besides, the function $f$ is determined by the data processing method, and the choice of a specific partially homomorphic cryptosystem should support the basic operation(s) in $f$. For example, the primitive of aggregate statistics~\cite{HE:Paillier:ccs2011} is addition, so the Paillier scheme~\cite{HE:Paillier99} can be the first choice; while the distance calculation~\cite{HE:BGN:ccs15} requires one more multiplication, thus, the BGN scheme~\cite{HE:BGN05} may be preferred. Furthermore, in Equation~(\ref{eq:dataprocessing_c}), $\vec{D}_0$ is the data consumer's ciphertext vector, and $\vec{D}_{c_i}$ indicates that the data contributor $o_{c_i}$ is one of the $m$ valid data contributors. More precisely, $m$ is the size of whitelist on the certificated bulletin board, and its default value is $n$. However, if either of the two-layer batch verifications fails, $m$ will be updated in the following tracing and revocation phase. For brevity in notations, we use $\mathbb{C}$ to denote the indexes of $m$ valid data contributors, \ie, $\mathbb{C} = \{c_1, c_2, \ldots, c_m\}$.


Next, the service provider sends $R$ to the registration center for decryption. We note that the registration center can only perform decryption for acknowledged times, which should be publicly announced on the certificated bulletin board. For example, in the aggregate statistic over a valid dataset of size $m$, the registration center just needs to do one decryption, and cannot do more than required. The reason is that the service provider can still obtain the correct aggregate result by decrypting all $m$ encrypted raw data.

Upon getting the plaintext $\gamma$, the service provider can compare it with $\delta$, and obtain the comparison result $\vartheta$. For convenience, the concrete-value result $\gamma$ and the comparison result $\vartheta$ are collectively called \emph{outcome}. We note that the outcome may be in different formats, \eg, average speed in location-based aggregate statistic~\cite{HE:Paillier:ccs2011}, shopping suggestion in private recommendation~\cite{HE:paillier:PrivateRecommendation12TIFS}, and friending strategy in social networking~\cite{matching12INFOCOM}. We assume that the outcome involves $\phi$ candidate data contributors, and the subscripts of their pseudo identities are denoted as $\mathbb{I} = \left\{I_1, I_2, \cdots, I_\phi\right\}.$

After data processing, to further reduce communication overhead, the service provider can aggregate $\phi$ candidate signatures into one signature. In our scheme, the aggregate signature $\sigma = \prod_{i \in \mathbb{I}} \sigma_{i}.$ Then, the service provider sends the final tuple to the data consumer, including the data service \emph{outcome}, the aggregate signature $\sigma$, the index set $\mathbb{I}$, and $\phi$ candidate ciphertexts $\{\vec{D}_{i}|i\in \mathbb{I}\}$.



\vspace{0.2em}
\noindent$\blacktriangleright$ \textbf{Second-Layer Batch Verification}
\vspace{0.2em}

Similar to the first-layer batch verification, the data consumer can verify the legitimacies of $\phi$ candidate data sources by checking whether
\begin{align}
\nonumber
  &\hat{e}\left(\sigma, g_2\right)\\
 =\quad &\hat{e}\left(\prod_{i \in \mathbb{I}} \emph{PID}{_{i}^1}, P_{1}\right) \hat{e}\left(\prod_{i=1} H(\emph{PID}{_{i}^2})^{h(D{_{i}})}, P_{2}\right).\label{BV2}
\end{align}
Here, the pseudo identities on the right hand side of the above equation can be fetched from the certificated bulletin board according to the index set $\mathbb{I}$.



\vspace{0.2em}
\noindent$\blacktriangleright$ \textbf{Outcome Verification}
\vspace{0.2em}

The homomorphic properties also enable the data consumer to verify the truthfulness of data processing. Under the condition that the data consumer knows her plaintext $V$, all the cross terms involving $\vec{D}_0$ in Equation~(\ref{eq:dataprocessing_c}) can be evaluated through multiplication by a constant $V$. Hence, part of the most time-consuming homomorphic multiplications in the original data processing are no longer needed in outcome verification. Besides, if for correctness, the data consumer just needs to evaluate on the $\phi$ candidate ciphertexts. Of course, she reserves the right to require the service provider to send her the other $(m - \phi)$ valid ones, on which the completeness can be verified.

In fact, if $\phi$ or $m - \phi$ is too large, the data consumer can take the strategy of random sampling for verification, where the $m$ valid pseudo identities on the certificated bulletin board can be used for the sampling indexes. Random sampling is a tradeoff between security and efficiency, and we shall illustrate its feasibility in Section~\ref{sec:two:datamarkets} and Section~\ref{section:computation:overhead}.

\vspace{0.4em}
\noindent \textbf{Phase \uppercase\expandafter{\romannumeral5}: Tracing and Revocation}
\vspace{0.4em}

The two-layer batch verifications only hold when all the signatures are valid, and fail even when there is a single invalid signature. In practice, a signature batch may contain invalid one(s) caused by accidental data corruption or possibly malicious activities launched by an external attacker. Traditional batch verifier would reject the entire batch, even if there is a single invalid signature, and thus waste the other valid data items. Therefore, tracing and/or recollecting invalid data items and their corresponding signatures are important in practice. If the second-layer batch verification fails, the data consumer can require the service provider to find out the invalid signature(s). Similarly, if the first-layer batch verification fails, the service provider has to find out the invalid one(s) by herself.



\begin{algorithm}[!t]                      
\caption{\textsc{$\ell$-depth-tracing}}          
\label{kdt}                           
\small
\begin{algorithmic}[1]
\Ensure{$S = \{\sigma_1, \cdots, \sigma_n\}$, $head = 1$, $tail = n$, $limit = \ell$, $whitelist = \varnothing,\ blacklist = \varnothing,\ resubmitlist = \varnothing$}
\Function{$\ell$-depth-Tracing}{$S, head, tail, limit$}
  \If{$|whitelist| + |blacklist| = n$ \textbf{or} $limit = 0$}
     \State \Return 
  \ElsIf{\Call{check-valid}{$S, head, tail$} = true}
    \State \Call{Add-to-whitelist}{$head, tail$}
  \ElsIf{$head = tail$} \Comment{\parbox[t]{.468\linewidth} {Single signature verification}}
     \State \Call{Add-to-blacklist}{$head, tail$}
  \Else \Comment{\parbox[t]{.79\linewidth}{Batch signatures verification from $\sigma_{head}$ to $\sigma_{tail}$}}
        \State $mid = \lfloor \frac{head + tail}{2} \rfloor$
        \State \Call{$\ell$-depth-tracing}{$S, head, mid, limit - 1$}
        \State \Call{$\ell$-depth-tracing}{$S, mid + 1, tail, limit - 1$}
  \EndIf
\EndFunction
\end{algorithmic}
\end{algorithm}
\normalsize

To extract invalid signatures, as shown in Algorithm~\ref{kdt}, we propose \textsc{$\ell$-depth-tracing} algorithm. We consider that the batch contains $n$ signatures. In addition, the whitelist, the blacklist, and the resubmit-list of pseudo identities are global variables, and are initialized as empty sets. If a batch verification fails, the service provider first finds out the mid-point as $mid = \lfloor \frac{1+n}{2} \rfloor$ (Line 9). Then, she performs batch verification on the first half ($head$ to $mid$) (Line 10) and the second half ($mid+1$ to $tail$) (Line 11), respectively. If either of these two halves causes a failure, the service provider repeats the same process on it. Otherwise, she adds the pseudo identities from the valid half to the whitelist (Line 4-5). The recursive process terminates, if validnesses of all the signatures have been identified or a pre-defined limit of search depth is reached (Line 2). A special case is the single signature verification, in which the service provider can determine its validness (Line 6-7). After this algorithm, the service provider can form the resubmit-list of pseudo identities by excluding those in the other two lists.



According to the blacklist on the certificated bulletin board, the registration center can reveal the real identities of those invalid data contributors. Given the data contributor $o_i$'s pseudo identity $\emph{PID}_i$, the registration center can use her master key $s_1$ to perform revealing by computing
\begin{align}\label{revocation}
   \emph{PID}{_i^2}\ \odot \emph{PID}{_i^1}^{s_1} = \emph{RID}_i \odot {P_0}^r \odot {g_1}^{s_1\cdot r} = \emph{RID}_i.
\end{align}
Upon getting a misbehaved data contributor's real identity, the registration center can revoke it from further usage if necessary, \eg, deleting her account from the online registration database. Thus, the revoked data contributor can no longer activate the tamper-proof device, which indicates that she does not have the right to submit data any more.

\section{Security Analysis}\label{securityanalysis}
In this section, we analyze the security of TPDM in terms of the desirable properties preconcerted in Section~\ref{adversary:security}.


\subsection{Data Authentication and Data Integrity}\label{proofauthentication}
Data authentication and data integrity are regarded as two basic security requirements in the data acquisition layer. The signature in TPDM $\sigma_i = SK{_i^1} \cdot SK{_i^2}^{h(D_i)}$ is actually a one-time identity-based signature. We now prove that if the Computational Diffie-Hellman (CDH) problem in the bilinear group $\mathbb{G}_1$ is hard~\cite{IBE2001}, an attacker cannot successfully forge a valid signature on behalf of any registered data contributor except with a negligible probability.


First, we consider \textsf{Game 1} between a challenger and an attacker as follows:
\begin{itemize}[leftmargin=.6cm]
\setlength{\itemindent}{-.4cm}
\item[] \textbf{Setup:} The challenger starts by giving the attacker the system parameters $g_1$ and $P_0$. The challenger also offers a pseudo identity $\emph{PID}_i = \langle \emph{PID}_i^1, \emph{PID}_i^2 \rangle$ to the attacker, which simulates the condition that the pseudo identities are posted on the certificated bulletin board in TPDM.

\item[] \textbf{Query:} We assume that the attacker does not know how to compute the MapToPoint hash function $H(\cdot)$ and the one-way hash function $h(\cdot)$. However, she can ask the challenger for the value $H(\emph{PID}{_i^2})$ and the one-way hashes $h(\cdot)$ for up to $n$ different messages.

\item[] \textbf{Challenge:} The challenger asks the attacker to pick two random messages $M_{i_1}$ and $M_{i_2}$, and to generate two corresponding signatures $\sigma_{i_1}$ and $\sigma_{i_2}$ on behalf of the data contributor $o_i$.

\item[] \textbf{Guess:} The attacker returns $\langle M_{i_1}, \sigma_{i_1} \rangle$ and $\langle M_{i_2}, \sigma_{i_2} \rangle$ to the challenger. We denote the attacker's advantage in winning \textsf{Game 1} to be
    \begin{align}
    \epsilon_1 = \text{Pr}[\sigma_{i_1}\ \text{and}\ \sigma_{i_2}\ \text{are valid}].
    \end{align}
\end{itemize}
We further claim that our signature scheme is adaptively secure against existential forgery, if $\epsilon_1$ is negligible. We prove our claim using \textsf{Game 2} by reduction~\cite{book:KatzLindell2014}.

Second, we assume that there exists a probabilistic polynomial-time algorithm $\mathcal{A}$ such that it has the same non-negligible advantage $\epsilon_1$ as the attacker in \textsf{Game 1}. Then, we will construct \textsf{Game 2}, in which an attacker $\mathcal{B}$ can make use of $\mathcal{A}$ to break the CDH assumption with non-negligible probability. In particular, $\mathcal{B}$ is given $(g_1, {g_1}^a, {g_1}^b, {g_1}^c, d)$ for unknown $(a, b, c)$ and known $d$, and is asked to compute ${g_1}^{2ab} \cdot {g_1}^{cd}$. We note that computing ${g_1}^{2ab} \cdot {g_1}^{cd}$ is as hard as computing ${g_1}^{ab}$, which is the original CDH problem. We present the details of \textsf{Game 2} as follows:

\begin{itemize}[leftmargin=.6cm]
\setlength{\itemindent}{-.4cm}
\item[] \textbf{Setup:} $\mathcal{B}$ makes up the parameters $g_1$ and $P_0 = {g_1}^a$, where $a$ plays the role of the master key $s_1$ in TPDM. Besides, $\mathcal{B}$ also provides $\mathcal{A}$ with a pseudo identity $\emph{PID}_i = \langle\emph{PID}_i^1, \emph{PID}_i^2\rangle = \langle {g_1}^b, \emph{RID}_i \odot {g_1}^{ab}\rangle$. Here, $b$ functions as the random nonce $r$ in TPDM.

\item[] \textbf{Query:} $\mathcal{A}$ then asks $\mathcal{B}$ for the value $H(\emph{PID}_i^2)^{s_2}$, and $\mathcal{B}$ replies with ${g_1}^c$. We note that $H(\emph{PID}_i^2)$ is the only MapToPoint hash operation needed to forge the data contributor $o_i$'s valid signatures. Besides, $\mathcal{A}$ picks $n$ random messages, and requests $\mathcal{B}$ for their one-way hash values $h(\cdot)$. $\mathcal{B}$ answers these queries using a random oracle: $\mathcal{B}$ maintains a table to store all the answers. Upon receiving a message, if the message has been queried before, $\mathcal{B}$ answers with the stored value; otherwise, she answers with a random value, which is stored into the table for later usage. Except for the $x$-th and $y$-th queries (\ie, messages $M_x$ and $M_y$), $\mathcal{B}$ answers with the values $d_1$ and $d_2$, respectively, where $d_1 + d_2 = d$.

\item[] \textbf{Challenge:} When the query phase is over, $\mathcal{B}$ asks $\mathcal{A}$ to choose two random messages $M_{i_1}$ and $M_{i_2}$, and to sign them on behalf of the data contributor $o_i$.

\item[] \textbf{Guess:} $\mathcal{A}$ returns two signatures $\sigma_{i_1}$ and $\sigma_{i_2}$ on the messages $M_{i_1}$ and $M_{i_2}$ to $\mathcal{B}$. We note that $M_{i_1}$ and $M_{i_2}$ must be within the $n$ queried messages; otherwise, $\mathcal{A}$ does not know $h(M_{i_1})$ and $h(M_{i_2})$. Furthermore, if $M_{i_1} = M_x$ and $M_{i_2} = M_y$ or $M_{i_1} = M_y$ and $M_{i_2} = M_x$, $\mathcal{B}$ then computes $\sigma_{i_1} \cdot \sigma_{i_2}$, which is equivalent to:
\begin{align}
\nonumber
  &SK{_i^1} \cdot SK{_i^2}^{h(M_{i_1})} \cdot SK{_i^1} \cdot SK{_i^2}^{h(M_{i_2})}\\
\nonumber
=\quad &{SK{_i^1}}^2 \cdot {SK{_i^2}}^{h(M_{i_1}) + h(M_{i_2})}\\
=\quad &{g_1}^{2ab} \cdot {g_1}^{cd}.
\end{align}
After obtaining $\sigma_{i_1} \cdot \sigma_{i_2}$, $\mathcal{B}$ solves the given CDH instance successfully. We note that $\mathcal{A}$'s advantage in breaking TPDM is $\epsilon_1$, and the probability that $\mathcal{A}$ picks $M_x$ and $M_y$ is $\frac{2}{n(n - 1)}$. Thus, the probability of $\mathcal{B}$'s success is:
\begin{equation}
\epsilon_2 = \text{Pr}[\mathcal{B}\ \text{succeeds}] = \frac{2\epsilon_1}{n(n - 1)}.
\end{equation}
\end{itemize}
Since $\epsilon_1$ is non-negligible, $\mathcal{B}$ can solve the CDH problem with the non-negligible probability $\epsilon_2$, which contradicts with the assumption that the CDH problem is hard. This completes our proof. Therefore, our signature scheme is adaptively secure under the random oracle model.

Last but not least, the first-layer batch verification scheme in TPDM is correct if and only if Equation~(\ref{BV1}) holds. By capitalizing the bilinear property of admissible pairing, the left hand side of Equation~(\ref{BV1}) expands as:
\begin{align}
\nonumber
 \quad &\hat{e}\left(\prod_{i=1}^{n} \sigma_i, g_2\right)\\
\nonumber
 = \quad &\hat{e}\left(\prod_{i=1}^{n}SK{_i^1}\cdot SK{_i^2}^{h(D_i)}, g_2\right)\\
\nonumber
 = \quad &\hat{e}\left(\prod_{i=1}^{n}SK{_i^1}, g_2\right) \hat{e}\left(\prod_{i=1}^{n}SK{_i^2}^{h(D_i)}, g_2\right)\\
\nonumber
 = \quad &\hat{e}\left(\prod_{i=1}^{n}\emph{PID}{_i^1}^{s_1}, g_2\right) \hat{e}\left(\prod_{i=1}^{n}H(\emph{PID}{_i^2})^{s_2h(D_i)}, g_2\right)\\
\nonumber
 = \quad &\hat{e}\left(\prod_{i=1}^{n}\emph{PID}{_i^1}, {g_2}^{s_1}\right) \hat{e}\left(\prod_{i=1}^{n}H(\emph{PID}{_i^2})^{h(D_i)}, {g_2}^{s_2}\right)\\
 = \quad &\hat{e}\left(\prod_{i=1}^{n}\emph{PID}{_i^1}, P_1\right) \hat{e}\left(\prod_{i=1}^{n}H(\emph{PID}{_i^2})^{h(D_i)}, P_2\right),\label{eq:pf:bv1}
\end{align}
which is the right hand side as required. 


In conclusion, our novel identity-based signature scheme is provably secure, and the properties of data authentication and data integrity are achieved.

\subsection{Truthfulness of Data Collection}\label{sec:bv2:proof}


To guarantee the truthfulness of data collection, we need to combat the partial data collection attack defined in Section~\ref{adversary:security}. We note that it is just a special case of \textsf{Game 1} in Section~\ref{proofauthentication}, where the service provider is the attacker. Hence, it is infeasible for the service provider to forge valid signatures on behalf of any registered data contributor. Such an appealing property prevents the service provider from injecting spurious data undetectably, and enforces her to truthfully collect real data.

Similar to data authentication and data integrity, the data consumer can verify the truthfulness of data collection by performing the second-layer batch verification with Equation~(\ref{BV2}). Proof of correctness is similar to Equation~(\ref{eq:pf:bv1}) for the first-layer batch verification, where we can just replace the aggregate signature $\sigma$ with $\prod_{i \in \mathbb{I}} \sigma_{i}$.


\subsection{Truthfulness of Data Processing}
We now analyze the truthfulness of data processing from two aspects, \ie, correctness and completeness.

\textbf{Correctness.} TPDM ensures the truthfulness of data collection, which is the premise of a correct data service. Then, given a truthfully collected dataset, the data consumer can evaluate the $\phi$ candidate data sources, which is consistent with the original data processing due to the homomorphic properties of the partially homomorphic cryptosystem. 


\textbf{Completeness.} In fact, our design provides the property of completeness by guaranteeing the correctness of $n$, $m$, and $\phi$, which are the numbers of total, valid, and candidate data contributors, respectively:

First, the service provider cannot deliberately omit a data contributor's real data. The reason is that if the data contributor has submitted her encrypted raw data, without finding her pseudo identity on the certificated bulletin board, she would obtain no reward for data contribution. Therefore, she has incentives to report data missing to the registration center, which in turn ensures the correctness of $n$.

Second, we consider that the service provider compromises the number of valid data contributors $m$ in two ways: one is to put a valid data contributor's pseudo identity into the blacklist; the other is to put an invalid pseudo identity into the whitelist. We discuss these two cases separately: 1) In the first case, the valid data contributor would not only receive no reward, but may also be revoked from the online registration database. Hence, she has strong incentives to resort to the registration center for arbitration. Besides, we claim that the service provider wins the arbitration except with negligible probability. We give the detailed proof via \textsf{Game 3} between a challenger and an attacker:
\begin{itemize}[leftmargin=.6cm]
\setlength{\itemindent}{-.4cm}
\item[] \textbf{Setup:} The challenger first gives the attacker $m$ valid data tuples, denoted as $\{\langle \emph{PID}_i, \vec{D}{_i}, \sigma_{i} \rangle|i \in \mathbb{C}\}$. This simulates the data submissions from $m$ valid data contributors.
\item[] \textbf{Challenge:} The challenger asks the attacker to pick a random data contributor $o_{i}$ within the $m$ given ones, and then requests the attacker to generate a signature $\sigma_i^*$ on the ciphertext vector $\vec{D}{_i}$.
\item[] \textbf{Guess:} The attacker returns $\sigma_i^*$ to the challenger. The attacker wins \textsf{Game 3}, if $\sigma_i^* \neq \sigma_i$, $\sigma_i^*$ passes the challenger's verification, and $\sigma_i$ fails in the verification.
\end{itemize}
Next, we demonstrate that the attacker's winning probability in \textsf{Game 3}, denoted as
\begin{equation}
\epsilon_3 = \text{Pr}[\sigma_i^* \neq \sigma_i,\ \sigma_i^*\ \text{passes verification, and}\ \sigma_i\ \text{fails}],
\end{equation}
is negligible. On one hand, the verification scheme in TPDM is publicly verifiable, which indicates that the challenger can verify the legitimacies of both $\sigma_i^*$ and $\sigma_i$ through checking whether
\begin{equation}\label{eq:single:verify}
\left\{
\begin{aligned}
&\hat{e}\left(\sigma_i^*, g_2\right) = \hat{e}\left(\emph{PID}{_i^1}, P_{1}\right) \hat{e}\left(H(\emph{PID}{_i^2})^{h(D_i)}, P_{2}\right),\\
&\hat{e}\left(\sigma_i, g_2\right) \neq \hat{e}\left(\emph{PID}{_i^1}, P_{1}\right) \hat{e}\left(H(\emph{PID}{_i^2})^{h(D_i)}, P_{2}\right),
\end{aligned}
\right.
\end{equation}
hold at the same time. We note that the above two equations conform to the formula of single signature verification, \ie, $n = 1$ in Equation~(\ref{BV1}). However, the second one contradicts with our assumption that $o_i$ is a valid data contributor. On the other hand, $\sigma_i^*$ passes the challenger's verification, while $\sigma_i^*$ is not equal to $\sigma_i$, which implies that $\sigma_i^*$ is a valid signature forged by the attacker. As shown in \textsf{Game 1}, the probability of successfully forging a valid signature $\epsilon_1$ is negligible, and thus the attacker's winning probability in \textsf{Game 3} $\epsilon_3$ is negligible as well. This completes our proof of thwarting the first case; 2) The second case is essentially the tracing and revocation phase in Section~\ref{DesignDetails}, where the batch of signatures contains invalid ones. Therefore, this case cannot pass two-layer batch verifications. Besides, the greedy service provider has no incentives to reward those invalid data contributors, which could in turn destabilize the data market. Joint considering above two cases, our scheme TPDM can guarantee the correctness of $m$.

Third, as stated in outcome verification, the data consumer reserves the right to verify over all $m$ valid data items, and the service provider cannot just process a subset without being found. Thus, the correctness of $\phi$ is assured.

In a nutshell, TPDM can guarantee the truthfulness of data processing in the data trading layer.





\subsection{Data Confidentiality}\label{sec:security:dataconf}
Considering the potential economic value and the sensitive information contained in raw data, data confidentiality is a necessity in data markets. Since partially homomorphic encryption provides semantic security (\eg,~\cite{HE:Paillier99,elgamal1985public,HE:BGN05}), by definition, except the registration center, any probabilistic polynomial-time adversary cannot reveal the contents of raw data. Moreover, although the registration center holds the private key, she cannot learn the sensitive raw data as well, since neither the service provider nor the data consumer directly forwards the original ciphertexts of the data contributors for decryption. Therefore, data confidentiality is achieved against all these system participants.


\subsection{Identity Preservation}
To protect a data contributor's unique identifier in data markets, her real identity is converted into a random pseudo identity. We note that the two parts of a pseudo identity are actually two items of an ElGamal-type ciphertext, which is semantically secure under the chosen plaintext attack~\cite{elgamal1985public}. Furthermore, the linkability between a data contributor's signatures does not exist, because the pseudo identities for different signing instances are indistinguishable. Hence, identity preservation can be ensured.

\subsection{Semi-honest Registration Center}\label{sec:sec:rc}
Registration center in TPDM performs two main tasks: one is to maintain the online database of legal registrations; the other is to set up the partially homomorphic cryptosystem.

First, as we have clarified in Section~\ref{sec:security:dataconf}, TPDM guarantees data confidentiality against the registration center. Thus, although she maintains the database of real identities, she cannot link them with corresponding raw data. Second, partially homomorphic encryption schemes (\eg, \cite{HE:Paillier99,elgamal1985public,HE:BGN05}) normally provide a proof of decryption, which indicates that the registration center cannot corrupt the decrypted results undetectably. Hence, she virtually has no effect on data processing and outcome verification. At last, we will further show the feasibility of distributing multiple registration centers in our evaluation part.

\section{Two Practical Data Markets}\label{sec:two:datamarkets}
In this section, from a practical standpoint, we consider two practical data markets, which provide fine-grained profile matching and multivariate Gaussian distribution fitting, respectively. The major difference between these two data markets is whether the data consumer has inputs.

\subsection{Fine-grained Profile Matching}\label{profilematching}

We first elaborate on a classic data service in social networking, \ie, fine-grained profile matching. Unlike the directly interactive scenario in~\cite{matching12INFOCOM}, our centralized data market breaks the limit of neighborhood finding. In particular, a data consumer's friending strategy can be derived from a large scale of data contributions. For convenience, we shall not differentiate ``profile" from ``raw data" in the profile-matching scenario considered here.

During the initial phase of profile matching, the service provider, \eg, Twitter or OkCupid, defines a public attribute vector consisting of $\beta$ attributes $\mathbf{A} = (A_1, A_2, \cdots, A_\beta)$, where $A_i$ corresponds to a personal interest, such as movie, sports, cooking, and so on. Then, to create a fine-grained personal profile, a data contributor $o_i$, \eg, a Twitter or OkCupid user, selects an integer $u_{ij} \in [0, \theta]$ to indicate her level of interest in $A_j \in \mathbf{A}$, and thus forms her profile vector $\vec{U}_i= ( u_{i1}, u_{i2}, \cdots, u_{i\beta}).$ Subsequently, the data contributor $o_i$ submits $\vec{U}_i$ to the service provider for matching process.


To facilitate profile matching, the data consumer also needs to provide her profile vector $\vec{V} = ( v_1, v_2, \cdots, v_\beta)$ and an acceptable similarity threshold $\delta$, where $\delta$ is a non-negative integer. Without loss of generality, we assume that the service provider employs \emph{Euclidean distance} $f(\cdot)$ 
to measure the similarity between the data contributor $o_i$ and the data consumer, where $f(\vec{U}_i, \vec{V}) = \sqrt{\sum_{j = 1}^{\beta}{(u_{ij} - v_j)}^2}$. We note that if $f(\vec{U}_i, \vec{V}) < \delta,$ then the data contributor $o_i$ is a matching target to the data consumer. In what follows, to simplify construction, we covert the matching metric $f(\vec{U}_i, \vec{V}) < \delta$ to its squared form $\sum_{j = 1}^{\beta}{(u_{ij} - v_j)}^2 < \delta^2.$

\subsubsection{Recap of Adversary Model}
\begin{figure}[!t]
\centering
\includegraphics [width = 0.96\columnwidth]{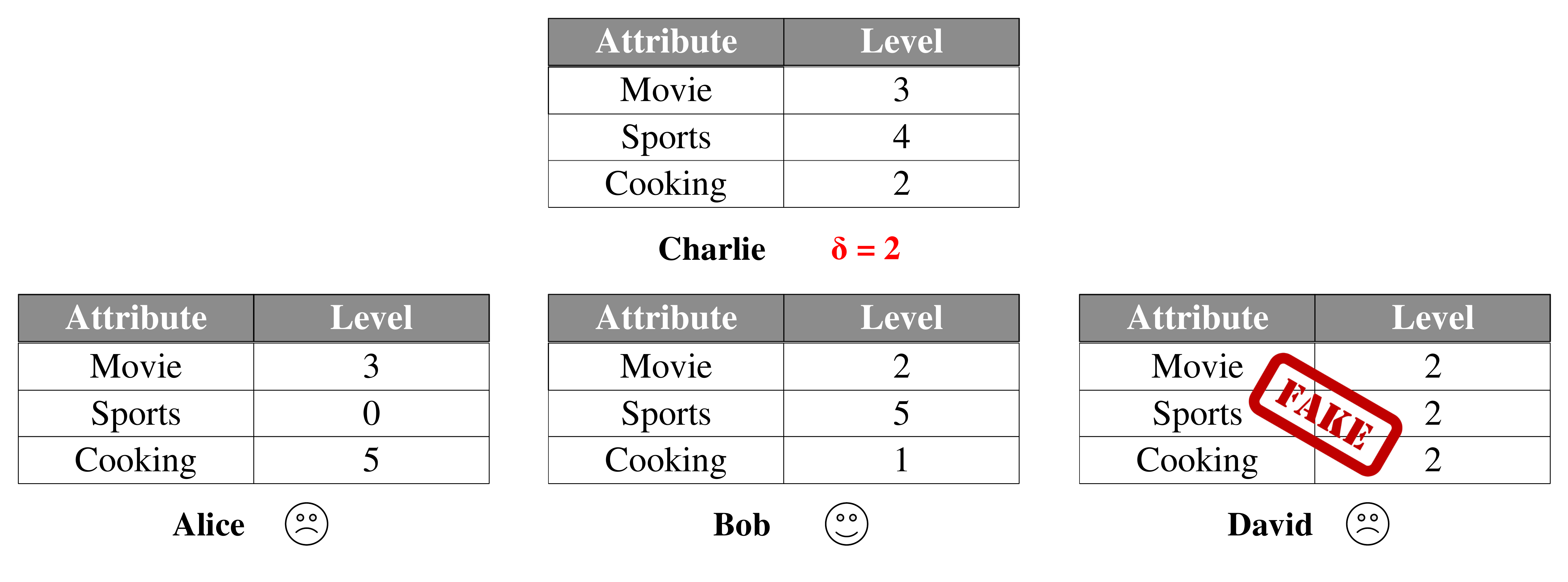}
\caption{An illustration of fine-grained profile matching.}
\label{fig:profilematchingexample}
\end{figure}

Before introducing our concrete construction, we first give a brief review of the adversary model and corresponding security requirements in the context of profile matching.

As shown in Fig.~\ref{fig:profilematchingexample}, Alice and Bob are registered data contributors, and Charlie is a data consumer. Here, the partial data collection attack means that to reduce data acquisition cost, the service provider may insert unregistered/fake David's profile. Besides, the partial data processing attack indicates that to reduce operation cost, the service provider may just evaluate the similarity between Charlie and Alice, while generating a random result for Bob. Moreover, the no data processing attack implies that the service provider just returns two random matching results without processing the profiles of both Alice and Bob.

Our joint security requirements of privacy preservation and data truthfulness mainly include two aspects: 1) Without leaking the real identities and the profiles of Alice and Bob, the service provider needs to prove the legitimacies of Alice and Bob to Charlie; 2) Without revealing Alice's and Bob's profiles, Charlie can verify the correctness and completeness of returned matching results.

\subsubsection{BGN-Based Construction}\label{sec:bgn:construction}
Given the profile-matching scenario considered here, we utilize a partially homomorphic encryption scheme based on bilinear maps, called Boneh-Goh-Nissim (BGN) cryptosystem~\cite{HE:BGN05}. This is because we only require the oblivious evaluation of quadratic polynomials, \ie, $\sum_{j = 1}^{\beta}{(u_{ij} - v_j)}^2$. In particular, the BGN scheme supports any number of homomorphic additions after a single homomorphic multiplication. Now, we briefly introduce how to adapt TPDM to this practical data market. Due to the limitation of space, we focus on the major phases, including data submission, data processing, and outcome verification.

\textbf{Data Submission:} When a data contributor $o_i$ intends to submit her profile $\vec{U}_i$, she employs the BGN scheme to do encryption, and gets the ciphertext vector:
\begin{align}
\vec{D}_i = \left(E(u_{ij}), E({u_{ij}}^2)\right)|_{j \in [1, \beta]}.
\end{align}
Afterwards, the data contributor $o_i$ computes the signature $\sigma_i$ on $\vec{D}_i$ using her secret key $SK_i$:
\begin{equation}\label{eq:pm:signature}
\sigma_i = SK{_i^1} \cdot SK{_i^2}^{h(D_i)},
\end{equation}
where $D_i = E(u_{i1}) \parallel \ldots \parallel E(u_{i\beta}) \parallel E({u_{i1}}^2)\parallel \ldots \parallel E({u_{i\beta}}^2),$ and ``$\parallel$" is a message concatenation operation.

By the end of a time period, $n$ distinct data contributors submit their tuples $\{\langle \emph{PID}_i, \vec{D}{_i}, \sigma_i \rangle|i\in [1, n]\}$ to the service provider, on which the first-layer batch verification can be conducted using Equation~(\ref{BV1}).

\textbf{Data Processing:} To facilitate generating a personalized friending strategy, the data consumer also needs to provide her encrypted profile vector $\vec{D}_0$ and a threshold $\delta$, where
\begin{align}\label{eq:cipherdc}
\vec{D}_0 = \left(E({v_j}^2), E(v_j)^{-2} = E(-2v_j)\right)|_{j \in [1, \beta]}.
\end{align}

Now, the service provider can directly do matching on the encrypted profiles. For brevity in expression, we assume that $o_i$ is one of the $m$ valid data contributors, \ie, $i\in\mathbb{C}$. Besides, to obliviously evaluate the similarity $f(\vec{U}_i, \vec{V})$, the service provider first preprocesses $\vec{D}_i$ and $\vec{D}_0$  by adding $E(1)$ to the first and the last places of two vectors, respectively, and obtains new vectors $\vec{C}_i = (C_{ij}^1, C_{ij}^2, C_{ij}^3)|_{j \in [1, \beta]}$ and $\vec{C}_0 = (C_{0j}^1, C_{0j}^2, C_{0j}^3)|_{j \in [1, \beta]}$, where
\begin{align}
\left( C_{ij}^1, C_{ij}^2, C_{ij}^3\right) &= \left(E(1), E(u_{ij}), E({u_{ij}}^2)\right),\label{eq:ci}\\
\left(C_{0j}^1, C_{0j}^2, C_{0j}^3\right) &= \left(E({v_j}^2), E(-2v_j), E(1)\right).\label{eq:c0}
\end{align}
After preprocessing, the service provider can compute the ``dot product'' of Equation~(\ref{eq:ci}) and Equation~(\ref{eq:c0}), by first applying homomorphic multiplication $\otimes$ and then homomorphic addition $\oplus$, and gets $R_{ij}$, where
\begin{align}\label{eq:square}
\nonumber
R_{ij} &= C_{ij}^1 \otimes C_{0j}^1 \oplus C_{ij}^2 \otimes C_{0j}^2 \oplus C_{ij}^3 \otimes C_{0j}^3\\
\nonumber
&= E\left({v_j}^2 + u_{ij}(-2v_j) + {u_{ij}}^2\right)\\
&= E\left((u_{ij} - v_j)^2\right).
\end{align}
Next, the service provider applies homomorphic additions $\oplus$ to $R_{ij}$ with $\forall j \in [1,\beta]$, and gets
\begin{align}
R_i = \underset{{j \in [1, \beta]}}{\oplus} E\left(\left(u_{ij} - v_j\right)^2\right) = E\left(\sum_{j = 1}^{\beta}{(u_{ij} - v_j)}^2\right).
\end{align}
We note that $R_{i}$ is actually an encryption of $f(\vec{U}_i, \vec{V})^2$, which indicates the similarity between the data contributor $o_i$ and the data consumer.


Then, the service provider sends $R_i$ to the registration center for decryption. We note that for each data contributor, the registration center just needs to do one decryption, \ie, supposing the size of whitelist on the certificated bulletin board is $m$, she can only perform $m$ decryptions in total. The registration center cannot do more decryptions than required, since the service provider may still obtain a correct and complete matching strategy by revealing the profiles of all the valid data contributors and the data consumer. However, this case requires at least $(m+1)\beta$ decryptions. Furthermore, to speed up BGN decryption in outcome verification, the registration center should retain the decrypted plaintexts in storage for a preset validity period.

When obtaining $f(\vec{U}_i, \vec{V})^2$, the service provider compares it with $\delta^2$, and thus can determine whether the data contributor $o_i$ matches the data consumer. We assume that $\phi$ data contributors are matched, and the subscripts of their pseudo identities are denoted as $\mathbb{I} = \{I_1, I_2, \cdots, I_\phi\}$.

After data processing, the service provider aggregates the signatures of $\phi$ matched data contributors into one signature. Then, she sends the aggregate signature, the indexes of $\phi$ matched data contributors, and their encrypted profile vectors to the data consumer, on which the second-layer batch verification can be performed with Equation~(\ref{BV2}). Besides, to prevent the service provider from changing/revaluating $(m-\phi)$ valid but unmatched data contributors in the completeness verification later, their similarities, \ie, $\{f(\vec{U}_i,\vec{V})^2|i\in\mathbb{C}, i\notin\mathbb{I}\},$ should also be forwarded. We note that the pseudo identities of $\phi$ matched data contributors can be viewed as the friending strategy, \ie, outcome in the general model, since the data consumer can resort to the registration center, as a relay, for handshaking with those matched data contributors.

\textbf{Outcome Verification:} During the validity period preset by the registration center, the data consumer can verify the truthfulness of data processing via homomorphic properties. For correctness, the data consumer just needs to evaluate the $\phi$ matched profiles. Of course, for completeness, the data consumer reserves the right to do verification over the other $(m-\phi)$ unmatched ones. We note that the data consumer, knowing her own profile vector $\vec{V}$, can compute Equation~(\ref{eq:square}) more efficiently through
\begin{align}
R_{ij} = E({v_j}^2) \oplus E\left(u_{ij}\right)^{-2v_j} \oplus E\left({u_{ij}}^2\right).
\end{align}
Thus, the most time-consuming homomorphic multiplications can be avoided in outcome verification. Moreover, we note that the registration center does not need to do decryption as in data processing, since she can just search a smaller-size table of plaintexts in the storage. If there is no matched one, the outcome verification fails, and the service provider will be questioned by the data consumer.

To further reduce verification cost, the data consumer can take the stratified sampling strategy in practice, \eg, in our evaluation on a real-world ratings dataset, for correctness, she may check all the $\phi$ matched data contributors, accounting for $4.49\%$ of the total 10000 samples, while only checking $0.27\%$ of the unmatched ones for completeness. In particular, regarding completeness verification, the data consumer can randomly choose part of the $(m-\phi)$ valid but unmatched data contributors, and then request the service provider to send her their aggregate signature and encrypted profile vectors for the second-layer batch verification and the outcome verification. Here, we assume that the greedy service provider cheats by not evaluating each data contributor in the original data processing with a probability $p$. Then, the probability of successfully detecting an attempt to return an incorrect/incomplete result, $\epsilon$, increases exponentially with the number of checks $c$, \ie, $\epsilon = 1 - (1 - p)^c$. For example, when $p = 20\%$ and $c = 10$, the success rate $\epsilon$ is already $90\%$. In fact, a concrete sampling strategy should depend on practical $\phi$, $m$, and $p$.

\subsection{Multivariate Gaussian Distribution Fitting}
We further consider a different data market, where the service provider captures the underlying probability distribution over the collected dataset, and offers such a distribution as a data service to the data consumer~\cite{jour:jsac2017:zheng,proc:mobihoc:zheng:gaussian}. This data service is called probability distribution fitting. For example, a data analyst, as the data consumer, may want to learn the distribution of residential energy consumptions.


Due to central limit theorem, we assume that the multivariate Gaussian distribution can closely approximate the raw data, which is a widely used assumption in statistical learning algorithms~\cite{book:06:Bishop}. For convenience, we continue to use the notations in profile matching, \ie, the attribute vector $\mathbf{A}$ now represents a vector of $\beta$ random variables. In particular, $\mathbf{A} \sim \mathcal{N}(\vec{\mu}, \mathbf{\Sigma})$, where $\vec{\mu}$ is a $\beta$-dimensional mean vector, and $\mathbf{\Sigma}$ is a $\beta\times\beta$ covariance matrix. Besides, the covariance matrix can be computed by:
\begin{align}
\mathbf{\Sigma} = \mathbb{E}\left[\mathbf{A}\mathbf{A}^T\right] - \vec{\mu}\vec{\mu}^T.
\end{align}
Here, $\mathbb{E}[\cdot]$ denotes taking expectation. We below focus on the key designs different from profile matching.

For data submission, the cipertext vector of the data contributor $o_i$ is changed into:
\begin{align}
\vec{D}_i = \left(E(u_{ij}), E({u_{ij}} \times {u_{ik}})\right)|_{j \in [1, \beta], k \in [j, \beta]},
\end{align}
where its first element is to facilitate computing the mean vector $\vec{\mu}$, while its second element is to help the service provider in evaluating the matrix $\mathbb{E}[\mathbf{A}\mathbf{A}^T]$ more efficiently. 

For data processing, the service provider first employs homomorphic additions $\oplus$ to obliviously compute the mean vector $\vec{\mu}$, where the ciphertext of its $j$-th element multiplying the number of valid data contributors $m$ is:
\begin{align}
\underset{{i \in \mathbb{C}}}{\oplus}E\left(u_{ij}\right) = E\left(\sum_{i \in \mathbb{C}}u_{ij}\right) = E(m \times \mu_j).
\end{align}
Additionally, to derive the covariance matrix $\mathbf{\Sigma}$, it suffices for the service provider to get $\mathbb{E}[\mathbf{A}\mathbf{A}^T]$. Here, the service provider can avoid the time-consuming homomorphic multiplications. For example, the $j$-th row, $k$-th column entry of $\mathbb{E}[\mathbf{A}\mathbf{A}^T]$, denoted by $\mathbb{E}[\mathbf{A}\mathbf{A}^T]_{jk}$, can be computed through:
\begin{align}
\nonumber
\underset{{i \in \mathbb{C}}}{\oplus} \left( E\left(u_{ij} \times u_{ik}\right)\right) &= E\left(\sum_{i \in \mathbb{C}} u_{ij} \times u_{ik}\right)\\
&= E\left(m \times \mathbb{E}\left[\mathbf{A}\mathbf{A}^T\right]_{jk}\right).\label{eq:AAT}
\end{align}
However, supposing that the data contributor $o_i$ excluded $\{E(u_{ij} \times u_{ik})|j\in[1, \beta], k \in [j, \beta]\}$ from her ciphertext vector $\vec{D}_i$, the service provider would need to perform $\frac{\beta(\beta+1)}{2}$ expensive homomorphic multiplications for the data contributor $o_i$ instead, because $E(u_{ij} \times u_{ik})$ in Equation~(\ref{eq:AAT}) needs to be derived by means of $E(u_{ij}) \otimes E(u_{ik})$.

For outcome verification, the data consumer can take the stratified random sampling strategy from two aspects: 1) She can randomly check parts of the mean vector $\vec{\mu}$ and the covariance matrix $\mathbf{\Sigma}$; 2) She can reevaluate a random subset of $m$ valid data items, and compare the new distribution with the returned distribution. If their distance is within a threshold, the data consumer would accept this outcome; otherwise, she rejects. We note that in the first case, the valid data contributors may need to re-sign those involved ciphertexts for the second-layer batch verification.

\section{Evaluation Results}\label{evaluation}

In this section, we show the evaluation results of TPDM in terms of computation overhead and communication overhead. We also demonstrate the feasibility of the registration center and the \textsc{$\ell$-depth-tracing} algorithm. 

\textbf{Datasets:} We use two real-world datasets, called R1-Yahoo! Music User Ratings of Musical Artists Version 1.0~\cite{Yahoo:WebScopeDatasets} and 2009 Residential Energy Consumption Survey (RECS) dataset~\cite{link:recs2009}, for the profile matching service and the distribution fitting service, respectively.

First, the Yahoo! dataset represents a snapshot of Yahoo! Music community's preference for various musical artists. It contains 11,557,943 ratings of 98,211 artists given by 1,948,882 anonymous users, and was gathered over the course of one month prior to March 2004. For profile matching, we choose $\beta$ common artists as the attributes, append each user's corresponding ratings ranging from 0 to 10, and thus form her fine-grained profile. Second, the RECS dataset, which was released by U.S. Energy Information Administration (EIA) in January 2013, provides detailed information about diverse energy usages in U.S. homes. This dataset was collected from 12,083 randomly selected households between July 2009 and December 2012. For distribution fitting, we view $\beta$ types of energy consumptions, \eg, electricity, natural gas, space heating, and water heating, as $\beta$ random variables, and intend to obtain the multivariate Gaussian distribution.



\textbf{Evaluation Settings:} We implemented TPDM using the latest Pairing-Based Cryptography (PBC) library~\cite{PBC}. The elliptic curves utilized in our identity-based signature scheme include a supersingular curve with a base field size of 512 bits and an embedding degree of 2, and a MNT curve with a base field size of 159 bits and an embedding degree of 6. In addition, the group order $q$ is 160-bit long, and all hashings are implemented in SHA1, considering its digest size closely matches the order of $\mathbb{G}_1$. The BGN cryptosystem is realized using Type A1 pairing, in which the group order is a product of two 512-bit primes. The running environment is a standard 64-bit Ubuntu 14.04 Linux operation system on a desktop with Intel(R) Core(TM) $i5$ $3.10GHz$.

\begin{table}[!t]
\caption{Time Overheads of Key Operations.} \label{tab:CurvePara}
\centering
\resizebox{\columnwidth}{!}
{
\begin{tabular}[t]{l|ccccc}
\hline
Curve & $\mathcal{R}(\mathbb{G}_1)$ & Pairing & MapToPoint & Exponentiation\\
\hline\hline
SS512 & 512 bits  & 0.999ms & 3.203ms & 1.179ms\\
MNT159 & 160 bits & 3.102ms & 0.029ms & 0.413ms\\
\hline
\end{tabular}
}
\end{table}

\textbf{Overheads of Key Operations:} Table~\ref{tab:CurvePara} presents the curve choices along with the computation time of key operations, where SS512 and MNT159 are abbreviated from the settings of the supersingular curve and the MNT curve in the identity-based signature scheme, respectively. $\mathcal{R}(\cdot)$ denotes the number of bits needed to optimally represent a group element. Besides, all the computation time of key operations is derived from the average of 10000 runs. 



\subsection{Computation Overhead}\label{section:computation:overhead}
\begin{figure*}[t]
\centering
\subfloat[Profile Matching]{\label{fig:EncryptionDE}
\includegraphics[width=0.68\columnwidth]{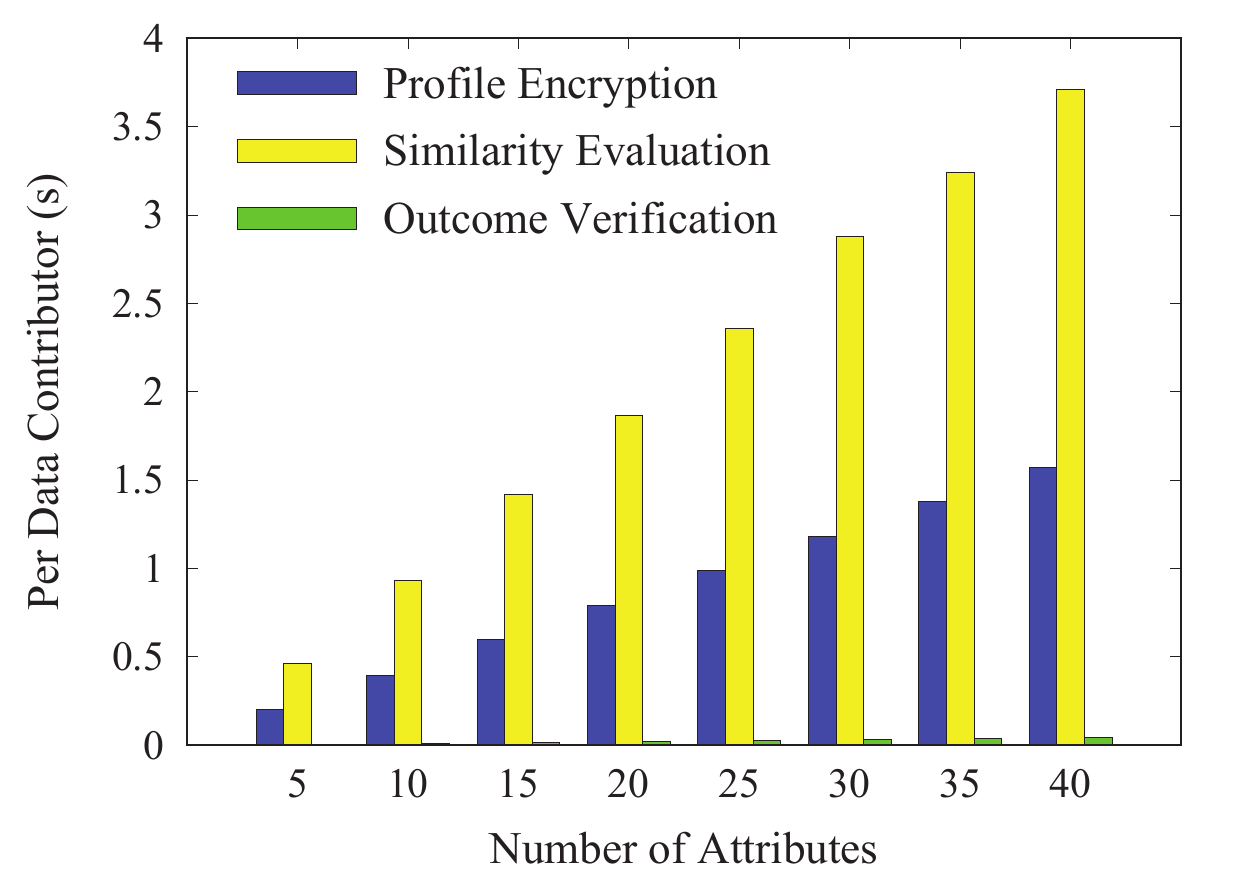}} 
\subfloat[Distribution Fitting]{\label{fig:distri}
\includegraphics[width=0.68\columnwidth]{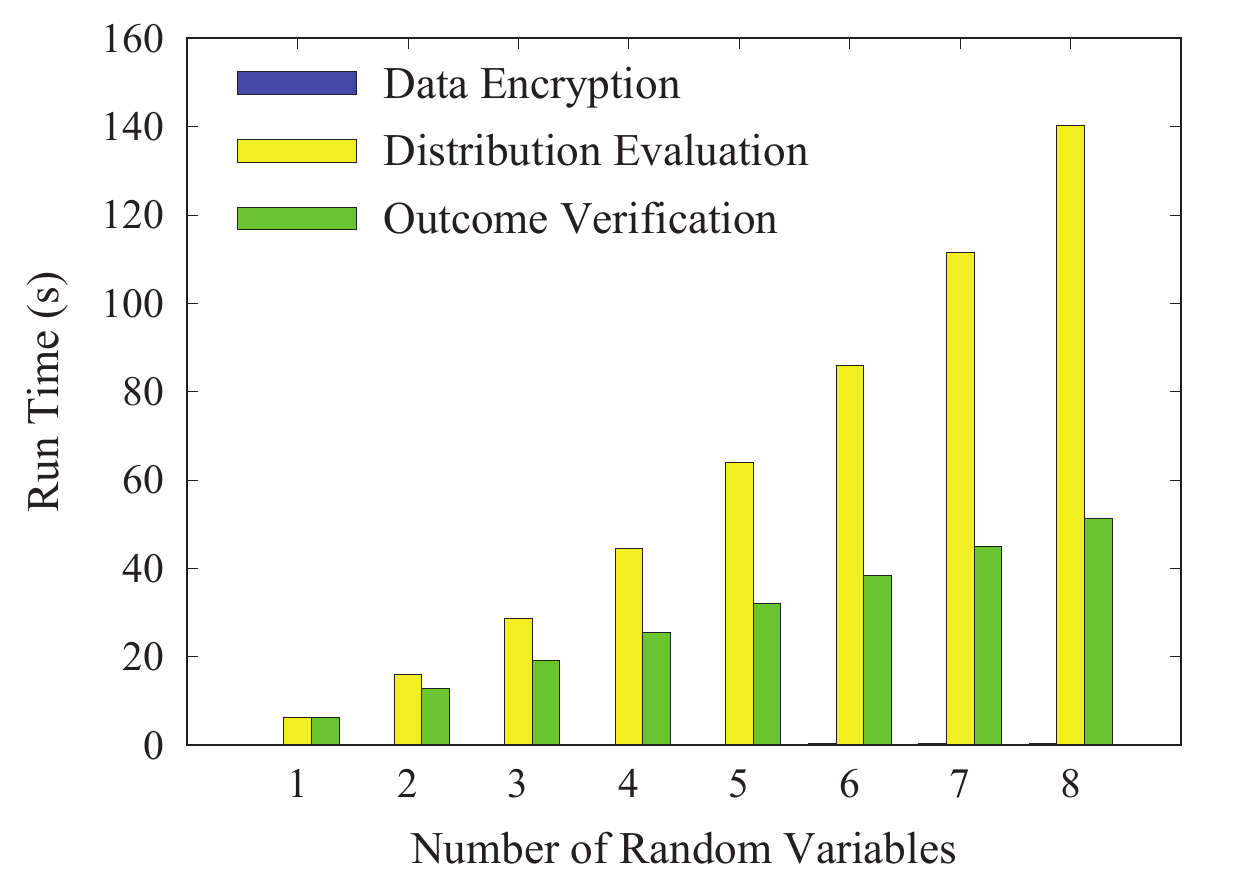}}
\subfloat[Batch Verification]{\label{fig:IBV}
\includegraphics[width=0.68\columnwidth]{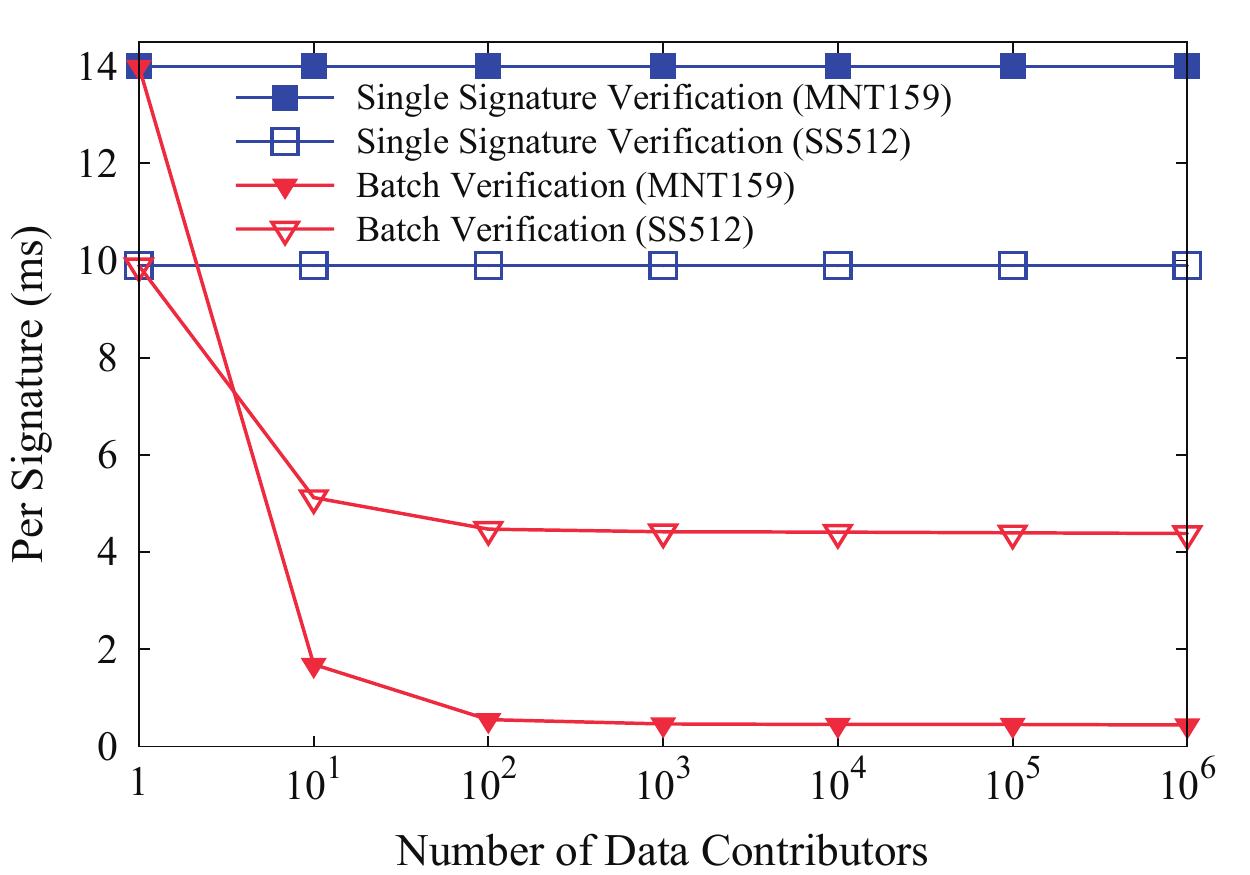}}
\caption{Computation overhead of TPDM.}
\end{figure*}

We show the computation overheads of four important components in TPDM, namely profile matching, distribution fitting, identity-based signature, and batch verification.

\textbf{Profile Matching:} In Fig.~\ref{fig:EncryptionDE}, we plot the computation overheads of profile encryption, similarity evaluation, and outcome verification per data contributor, when the number of attributes $\beta$ increases from 5 to 40 with a step of 5. From Fig.~\ref{fig:EncryptionDE}, we can see that the computation overheads of these three phases increase linearly with $\beta$. This is because the profile encryption requires $2\beta$ BGN encryptions, the similarity evaluation consists of $3\beta$ homomorphic multiplications and additions, and the outcome verification is composed of $3\beta$ homomorphic additions and $\beta$ exponentiations, which are both proportional to $\beta$. In addition, the outcome verification is light-weight, whose overhead per data contributor is only $1.17\%$ of the similarity evaluation's cost. Moreover, when $\beta = 10$, one decryption overhead at the registration center is 1.648ms in the original data processing, while in outcome verification, it is in tens of microseconds.



We further show the feasibility of the stratified sampling strategy in outcome verification. We analyze the matching ratio based on Yahoo! Music ratings dataset. Given $\beta = 10$, when a data consumer sets her threshold $\delta = 12$, she is matched with $4.49\%$ in average of the 10000 data contributors, who are selected randomly from the dataset. The relatively small matching ratio means that even if all matched data contributors are verified for correctness, it only incurs an overhead of 4.859s at the data consumer, which is roughly $0.05\%$ of the data processing workload at the service provider. Next, we simulate the partial data processing attack by randomly corrupting $20\%$ of unmatched data contributors, \ie, replacing their similarities with random values. Then, the data consumer can detect such type attack using 26 random checks in average for completeness, which incurs an additional overhead of 0.281s.


\textbf{Distribution Fitting:} Fig.~\ref{fig:distri} plots the computation overhead of the distribution fitting service, where the number of random variables $\beta$ increases from 1 to 8, and the number of valid data contributors $m$ is fixed at 10000. Besides, for outcome verification, the data consumer checks all the elements in the mean vector, while only checks the diagonal elements in the covariance matrix. From Fig.~\ref{fig:distri}, we can see that the computation overheads of the first two phases increase quadratically with $\beta$, whereas the computation overhead of the last phase increases linearly with $\beta$. The reason is that the data encryption phase consists of $\frac{\beta(\beta + 3)}{2}$ BGN encryptions for each data contributor, and the distribution evaluation phase mainly comprises $\frac{m\beta(\beta + 3)}{2}$ homomorphic additions. In contrast, the outcome verification phase mainly requires $2m\beta$ homomorphic additions. Furthermore, when $\beta = 8$, these three phases consume 0.402s, 140.395s, and 51.200s, respectively.

Jointly summarizing above evaluation results, TPDM performs well in both kinds of data markets. Therefore, the generality of TPDM can be validated.




\begin{table}[!t]
\caption{Computation Overhead of Identity-Based Signature Scheme.} \label{tab:IBS} 
\centering
\resizebox{\columnwidth}{!}{
\begin{tabular}[t]{l|m{2.2cm}<{\centering}m{2.2cm}<{\centering}m{2.2cm}<{\centering}}
\hline
 & \multicolumn{2}{c}{Preparation} & Operation\\
\hline
\hline
Setting& Pseudo Identity Generation & Secret Key Generation & Signing\\
SS512   & 4.698ms (39.40\%)    & 6.023ms (50.53\%) & 1.201ms (10.07\%)\\
MNT159 & 1.958ms (57.33\%) & 1.028ms (30.10\%)  & 0.429ms (12.57\%)\\
\hline
\end{tabular}
}
\end{table}

\textbf{Identity-Based Signature:} We now investigate the computation overhead of the identity-based signature scheme, including preparation and operation phases. In this set of simulations, we set the number of data contributors to be $10000$. Table~\ref{tab:IBS} lists the average time overhead per data contributor. From Table~\ref{tab:IBS}, we can see that the time cost of the preparation phase dominates the total overhead in both SS512 and MNT159. This outcome stems from that the pseudo identity generation employs ElGamal encryption, and the secret key generation is composed of one MapToPoint hash operation and two exponentiations. In contrast, the operation phase mainly consists of one exponentiation.

The above results demonstrate that the identity-based signature scheme in TPDM is efficient enough, and can be applied to the data contributors with mobile devices.




\textbf{Batch Verification:} To examine the efficiency of batch verification, we vary the number of data contributors from 1 to 1 million by exponential growth. The performance of the corresponding single signature verification is provided as a baseline. Fig.~\ref{fig:IBV} depicts the evaluation results using SS512 and MNT159, where verification time per signature (abbreviated as VTPS) is computed by dividing the total verification time by the number of data contributors. In particular, such a performance measure in an average sense can be found in~\cite{proc:infocom2010:batch:audit}. From Fig.~\ref{fig:IBV}, we can see that when the scale of data acquisition or data trading is small, \eg, when the number of data contributors is 10, TPDM saves 48.22$\%$ and 87.94$\%$ of VTPS in SS512 and MNT159, respectively. When the scale becomes larger, TPDM's advantage over the baseline is more remarkable. This is owing to the fact that TPDM amortizes the overhead of 3 time-consuming pairing operations among all the data contributors.



We now compare the batch verification efficiency of two settings. Although the baseline of MNT159 increases $41.44\%$ verification latency than that of SS512, MNT159's implementation is more efficient when the number of data contributors is larger than 10, \eg, when supporting as many as 1 million data contributors, MNT159 reduces $89.93\%$ of VTPS than SS512. We explain the reason by analyzing the asymptotic value of VTPS:
\begin{align}
\lim_{n \to +\infty} \frac{3 T_{par} + n T_{mtp} + n T_{exp}}{n} = T_{mtp} + T_{exp}.\label{eq:asymptotic:VTPS}
\end{align}
Here, we let $T_{par}$, $T_{mtp}$, and $T_{exp}$ denote the time overheads of a pairing operation, a MapToPoint hashing, and an exponentiation in Table~\ref{tab:CurvePara}, respectively. From Equation~(\ref{eq:asymptotic:VTPS}), we can draw that if the time overheads of additional operations, \eg, $T_{mtp}$ and $T_{exp}$, are approaching or even greater than that of pairing operation (\eg, in SS512), their effect cannot be elided. Besides, the expensive additional operations will cancel part of the advantage gained by batch verification. Even so, the batch verification scheme can still sharply reduce per-signature verification cost.

These evaluation results reveal that TPDM can indeed help to reduce the computation overheads of the service provider and the data consumer by introducing two-layer batch verifications, especially in large-scale data markets.

\subsection{Communication Overhead}\label{section:communication:overhead}
\begin{figure*}[t]
\centering
    \begin{minipage}{0.65\columnwidth}
        \centering
        \includegraphics[width=0.98\columnwidth]{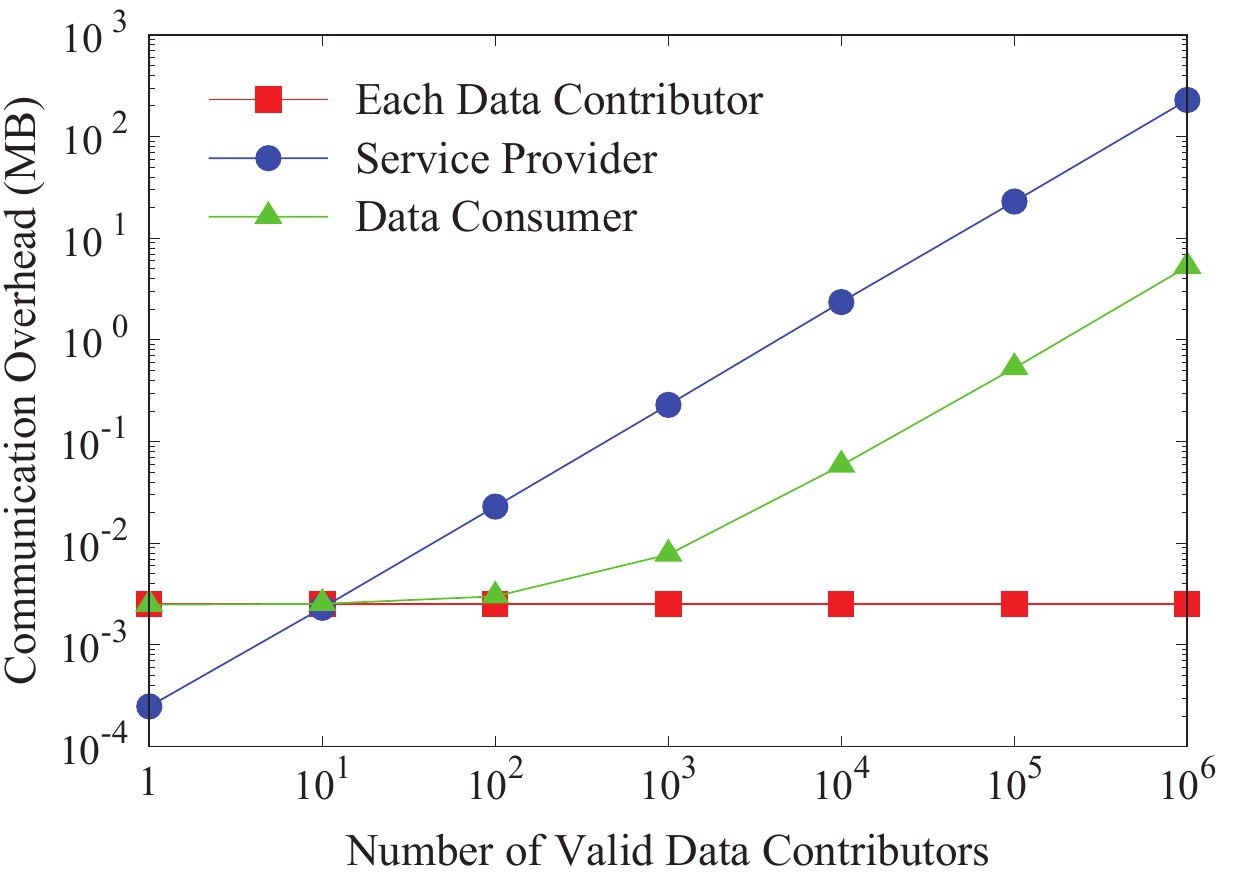} 
        \caption{Comm. overhead of profile matching.}\label{fig:co}
    \end{minipage}\hfill
    \begin{minipage}{0.65\columnwidth}
        \centering
        \includegraphics[width=0.98\columnwidth]{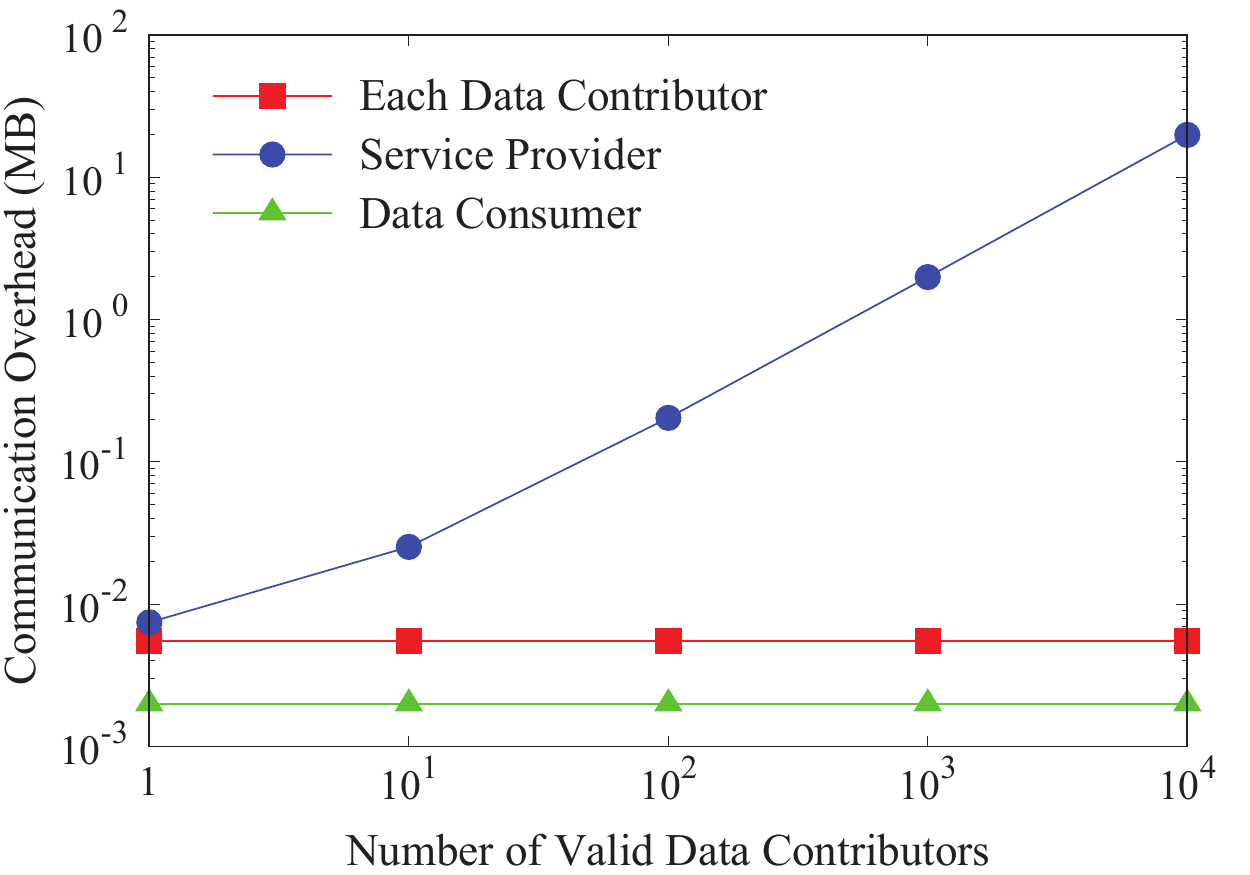}
        \caption{Comm. overhead of distribution fitting.}\label{fig:co:distri}
    \end{minipage}\hfill
    \begin{minipage}{0.65\columnwidth}
        \centering
        \includegraphics[width=0.98\columnwidth]{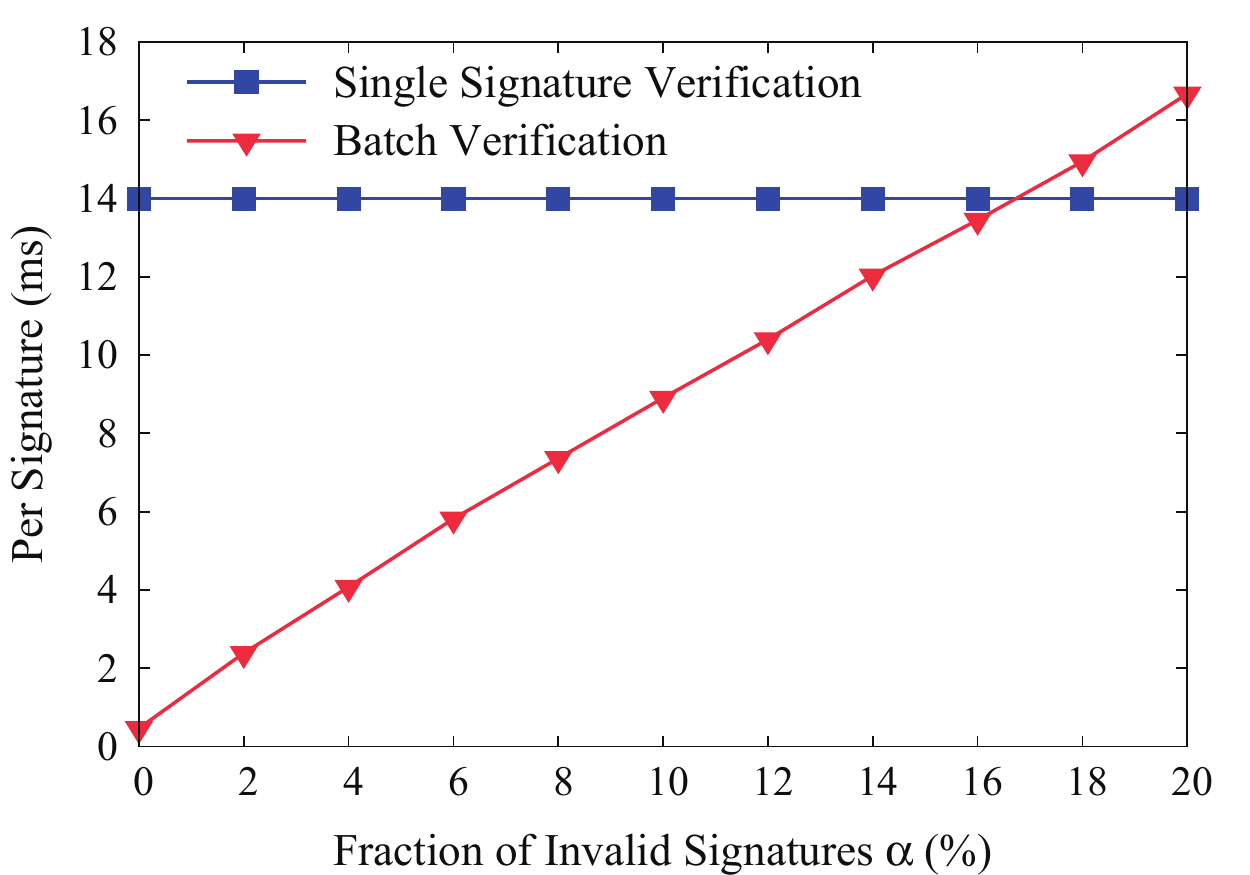}
        \caption{Feasibility of tracing algorithm.}\label{fig:KDT}
    \end{minipage}
\end{figure*}

In this section, we show the communication overheads of the profile matching service and the distribution fitting service separately.

Fig.~\ref{fig:co} plots the communication overhead of profile matching, where the identity-based signature scheme is implemented in MNT159, the number of attributes $\beta$ is fixed at 10, and the threshold $\delta$ takes $12$. Here, the communication overhead merely counts in the amount of sending content. Besides, we only consider the correctness verification. In fact, when the number of valid data contributors $m$ is $10^4$, if we check 26 unmatched ones for completeness, it incurs additional communication overheads of 80.03KB at the service provider, and 3.35KB at the data consumer. Moreover, our statistics on the dataset show a linear correlation between the numbers of matched data contributors $\phi$ and valid ones $m$, where the matching ratio is $4.24\%$ in average.

The first observation from Fig.~\ref{fig:co} is that the communication overheads of the service provider and the data consumer grow linearly with the number of valid data contributors $m$, while the communication overhead of each data contributor remains unchanged. The reason is that each data contributor just needs to do one profile submission, and thus its cost is independent of $m$. However, the service provider primarily needs to send $m$ encrypted similarities for decryption, and to forward the indexes and the ciphertexts of $\phi$ matched data contributors for verifications. Regarding the data consumer, her communication overhead mainly comes from one data submission and the delivery of $\phi$ encrypted similarities for decryption. These imply that the communication overheads of the service provider and the data consumer are both linear with $m$. Here, we note that x, y axes in Fig.~\ref{fig:co} are log-scaled, and thus the communication overhead of the data consumer, containing a constant of one data submission overhead, seems non-linear. In particular, when $m \leq 100$, one data submission overhead dominates the total communication overhead, and this interval looks like a horizontal line; while $m \geq 1000$, the communication overhead of delivering $\phi$ encrypted similarities dominates, and that interval appears linear.


The second key observation from Fig.~\ref{fig:co} is that when the number of valid data contributors $m = 10$, all the three system participants spend roughly the same network bandwidth. The cause lies in that the small matching ratio implies a small number of matched data contributors involved in the correctness verification. Specifically, when $m = 10$, the average number of matched data contributors is only about $0.4 < 1$, and the communication overheads of each data contributor, the service provider, and the data consumer are 2.60KB, 2.37KB, and 2.59KB, respectively.

We plot the communication overhead of multivariate Gaussian distribution fitting in Fig.~\ref{fig:co:distri}, where the number of random variables $\beta$ is set to be 8. From Fig.~\ref{fig:co:distri}, we can see that the communication overhead of the service provider increases linearly with the number of valid data contributors $m$. This is because the service provider mainly needs to send $2\beta m$ BGN-type ciphertexts for verifications, which is linear with $m$. By comparison, besides each data contributor, the data consumer's bandwidth overhead stays the same, since she needs to deliver $2\beta$ BGN-type ciphertexts for decryption, which is independent of $m$.

We finally note that the transmission of BGN-type ciphertexts dominates the total communication overheads in both data services, while the network overhead incurred by sending the pseudo identities and the aggregate signature is comparatively low. Therefore, we do not plot the cases for SS512, which are similar to Fig.~\ref{fig:co} and Fig.~\ref{fig:co:distri}. In particular, compared with MNT159, SS512 adds 132 bytes and 176 bytes at each data contributor in profile matching and distribution fitting, respectively. Besides, SS512 adds 44 bytes at the service provider in two data services, but incurs no extra bandwidth overhead at the data consumer.





\subsection{Feasibility of Registration Center}\label{eval:rc}
In this section, we consider the feasibility of the registration center from the perspectives of computation, communication, and storage overheads. We implement the identity-based signature scheme with MNT159. In addition, for profile matching, the number of attributes is fixed at 10, and the number of valid data contributors $m$ is set to be 10000. Accordingly, the number of matched ones $\phi$ is 449 at $\delta = 12$. For distribution fitting, we fix the number of random variables $\beta$ at 8, and set the number of valid data contributors to be 10000.

First, the primary responsibility of the registration center is to initialize the system parameters for the identity-based signature scheme and the BGN cryptosystem. Besides, she is required to perform totally $(m+\phi)$ and $\frac{(\beta +7)\beta}{2}$ decryptions in the profile matching service and the distribution fitting service, respectively. The total computation overheads are 16.692s and 3.065s in two data services, respectively, which are only $0.18\%$ and $2.11\%$ of the service provider's overall workloads. Furthermore, the one-time setup overhead can be amortized over several data services. Second, the main communication overheads of the registration center in two data services are incurred by returning decrypted results, which occupies the network bandwidth of 15.31KB and 0.23KB, respectively. Third, the storage overhead of the registration center mostly comes from maintaining the online database of registrations and the real-time certificated bulletin board, and caching the intermediate plaintexts. These two parts take up roughly 600.59KB and 586.11KB storage space in the profile matching service and the distribution fitting service, respectively.

%



In conclusion, our design of registration center has a light load, and can be implemented in a distributed manner, where each registration center can be responsible for one or a few data services. In particular, consistent hashing~\cite{STOC97:consistenthashing} can be employed to facilitate the information synchronization among multiple registration centers, \eg, guaranteeing a certain number of decryptions for each data service. Besides, using the standard techniques from~\cite{threshold:Pedersen1991}, the original partially homomorphic cryptosystems can be extended to their threshold multi-authority versions, which implies the improved robustness of TPDM by distributing several registration centers in data markets.




\subsection{Feasibility of Tracing Algorithm}

To evaluate the feasibility of \textsc{$\ell$-depth-tracing} algorithm when the batch verification fails, we generate a collection of 1024 valid signatures, and then randomly corrupt an $\alpha$-fraction of the batch by replacing them with random elements from the cyclic group $\mathbb{G}_1$. We repeat this evaluation with various values of $\alpha$ ranging from 0 to $20\%$, and compare verification time per signature (VTPS) in batch verification with that in single signature verification. Here, the overall batch verification latency includes the time cost spent in identifying invalid signatures. Fig.~\ref{fig:KDT} presents the evaluation results using the efficient MNT159. 


 As shown in Fig.~\ref{fig:KDT}, batch verification is preferable to single signature verification when the ratio of invalid signatures is up to $16\%$. The worst case of batch verification happens when the invalid signatures are distributed uniformly. In case the invalid signatures are clustered together, the performance of batch verification should be better. Furthermore, as shown in the initialization phase of Algorithm~\ref{kdt}, the service provider can preset a practical tracing depth, and let those unidentified data contributors do resubmissions.

\section{Related Work}\label{related_work}

In this section, we briefly review related work.

\subsection{Data Market Design}
%

In recent years, data market design has gained increasing interest from the database community. The seminal paper~\cite{DataMarket2011} by Balazinska~\et~discusses the implications of the emerging digital data markets, and lists the research opportunities in this direction. Koutris~\et~\cite{Koutris:2012:PODS}~presented a flexible data trading format, \ie, query-based data pricing. Later, Lin and Kifer~\cite{Lin:2014:VLDB} designed an arbitrage-free pricing function for arbitrary query formats. For personal data sharing, Li~\et~\cite{jour:cacm2017:li} proposed a theory of pricing private data based on differential privacy. Upadhyaya~\et~\cite{datalawyer:15sigmod} developed a middleware system, called DataLawyer, to formally specify data use policies, and to automatically enforce these pre-defined terms during data usage. Jung~\et~\cite{proc:infocom17:dataresale} focused on the dataset resale issue at the dishonest data consumers.

However, the original intention of above works is pricing data or monitoring data usage rather than integrating data truthfulness with privacy preservation in data markets, which is the consideration of this work\footnote{The early version of this paper~\cite{proc:icde17:niu} mainly focused on the profile matching service.}.

\subsection{Signcryption}
Data authentication and data confidentiality are two basic requirements in secure communication. To efficiently guarantee these two properties simultaneously, Zheng~\et~\cite{Zhengsigncryption} first introduced the terminology signcryption, which integrates digital signature with public key encryption. To reduce communication overhead, a number of identity-based signcryption schemes~\cite{BP2005IDBS,public_ciphertext}~were proposed. In most signcryption schemes (\eg,~\cite{BP2005IDBS,Zhengsigncryption}), a third party requires the knowledge of plaintext to verify the message's origin. In contrast to these works, Encrypt-then-Sign paradigm in~\cite{AN2002SE} and the identity-based signcryption scheme in~\cite{public_ciphertext} support public ciphertext authenticity, which convinces a third party (\eg, the data consumer in our model) of data sources' reliability without revealing the content of raw data.

Unfortunately, when existing signcryption schemes are directly applied to data markets, they only provide the truthfulness of data collection, but fail to support outcome verification. Besides, these schemes don't facilitate identity preservation and batch verification~\cite{jour:joc2012:bv}, which are necessary in practical data collection and data trading environments.

\subsection{Practical Computation on Encrypted Data}



To get a tradeoff between functionality and performance, partially homomorphic encryption (PHE) schemes were exploited to enable practical computation on encrypted data. Unlike those prohibitively slow fully homomorphic encryption (FHE) schemes~\cite{HE:FHE:STOC09,HE:FHE:BV14} that support arbitrary operations, PHE schemes focus on specific function(s), and achieve better performance in practice. A celebrated example is the Paillier cryptosystem~\cite{HE:Paillier99}, which preserves the group homomorphism of addition and allows multiplication by a constant. Thus, it can be utilized in data aggregation~\cite{HE:Paillier:ccs2011} and interactive personalized recommendation~\cite{matching12INFOCOM,HE:paillier:PrivateRecommendation12TIFS}. Yet, another one is ElGamal encryption~\cite{elgamal1985public}, which supports homomorphic multiplication, and it is widely employed in voting~\cite{HE:ElGamal:2014voting:helios:TPC}. Moreover, the BGN scheme~\cite{HE:BGN05} facilitates one extra multiplication followed by multiple additions, which in turn allows the oblivious evaluation of quadratic multivariate polynomials, \eg, shortest distance query~\cite{HE:BGN:ccs15} and optimal meeting location decision~\cite{HE:BGN:14TIFSopitmallocation}.

These schemes enable the service provider and the data consumer to efficiently perform data processing and outcome verification over encrypted data, respectively. Thus, they can remedy the potential defects in the conventional signcryption schemes. Additionally, we note that the outcome verification in data markets differs from the verifiable computation in outsourcing scenarios~\cite{proc:crypto11:private:verifiability}, since before data processing, the data consumer, as a client, does not hold a local copy of the collected dataset.

\section{Conclusion and Future Work}\label{conclusion}
In this paper, we have proposed the first efficient secure scheme TPDM for data markets, which simultaneously guarantees data truthfulness and privacy preservation. In TPDM, the data contributors have to truthfully submit their own data, but cannot impersonate others. Besides, the service provider is enforced to truthfully collect and process data. Furthermore, both the personally identifiable information and the sensitive raw data of data contributors are well protected. In addition, we have instantiated TPDM with two different data services, and extensively evaluated their performances on two real-world datasets. Evaluation results have demonstrated the scalability of TPDM in the context of large user base, especially from computation and communication overheads. At last, we have shown the feasibility of introducing the semi-honest registration center with detailed theoretical analysis and substantial evaluations.




As for further work in data markets, it would be interesting to consider diverse data services with more complex mathematic formulas, \eg, Machine Learning as a Service (MLaaS)~\cite{link:googelapi,link:aws:ml,link:new:azure,proc:usenix2016:mlaas,proc:icml16:cryptonet}. For a specific data service, it is well-motivated to uncover some novel security problems, such as privacy preservation and verifiability.

\appendices

\ifCLASSOPTIONcompsoc
  \section*{Acknowledgments}
\else
  \section*{Acknowledgment}
\fi

This work was supported in part by the State Key Development Program for Basic Research of China (973 project 2014CB340303), in part by China NSF grant 61672348, 61672353, 61422208, and 61472252, in part by Shanghai Science and Technology fund (15220721300, 17510740200), in part by the Scientific Research Foundation for the Returned Overseas Chinese Scholars, and in part by the CCF Tencent Open Research Fund (RAGR20170114). The opinions, findings, conclusions, and recommendations expressed in this paper are those of the authors and do not necessarily reflect the views of the funding agencies or the government. F. Wu is the corresponding author.
\ifCLASSOPTIONcaptionsoff
  \newpage
\fi



%

\bibliographystyle{IEEEtran}
\bibliography{TPDM_Ref,reflist-short,WVSI,HE,Futherwork,CCSTPC,Dataservice,latestwork}

%

\begin{IEEEbiography}[{\includegraphics[width=1in,height=1.25in,clip,keepaspectratio]{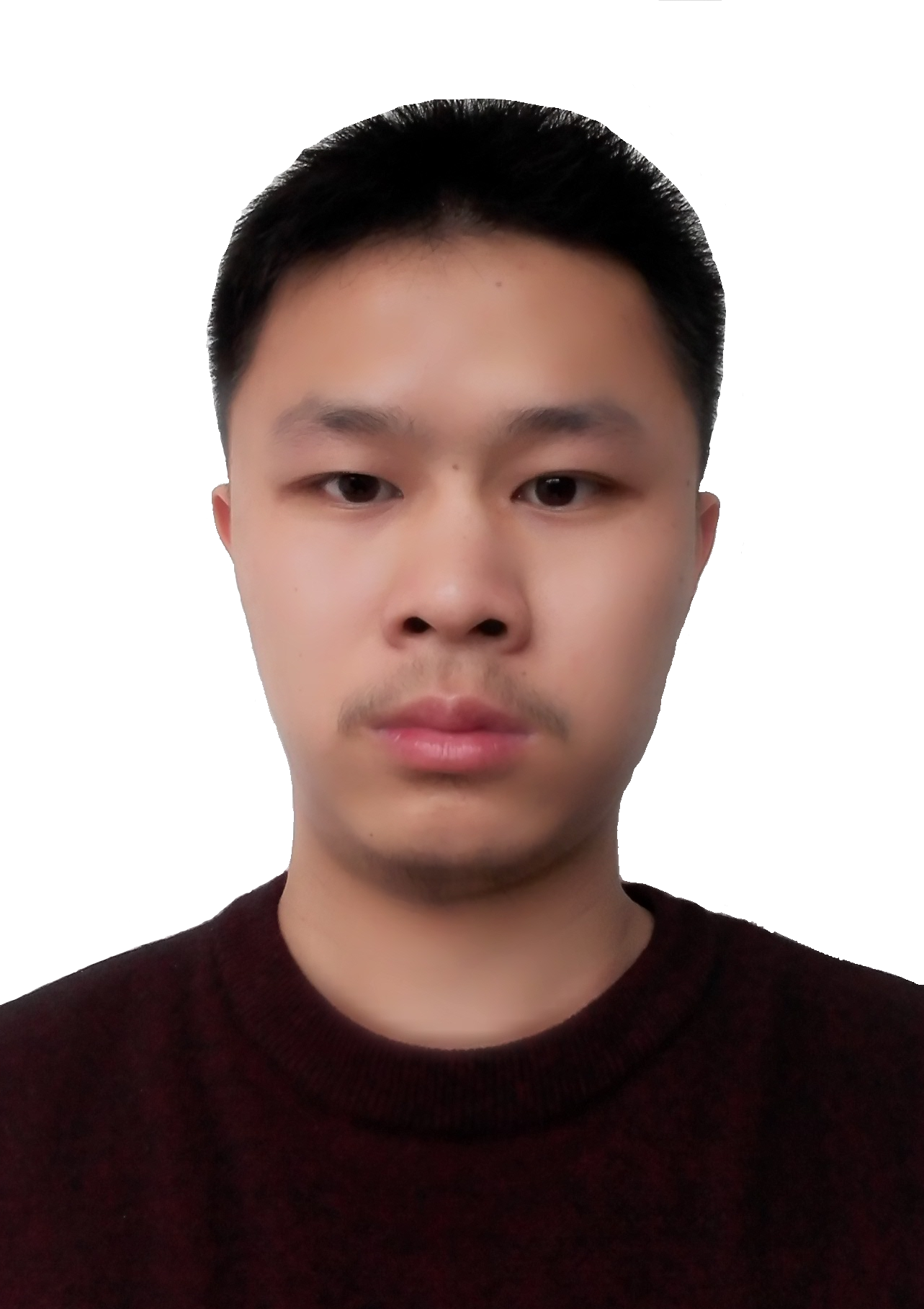}}]{Chaoyue Niu}
is a Ph.D. candidate from the Department of Computer Science and Engineering, Shanghai Jiao Tong University, P. R. China. His research interests include verifiable computation and privacy preservation in data management. He is a student member of ACM and IEEE.
\end{IEEEbiography}

\begin{IEEEbiography}[{\includegraphics[width=1in,height=1.25in,clip,keepaspectratio]{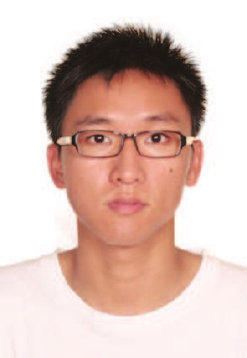}}]{Zhenzhe Zheng}
is a Ph.D. candidate from the Department of Computer Science and Engineering, Shanghai Jiao Tong University, P. R. China. His research interests include algorithmic game theory, resource management in wireless networking and data center. He is a student member of ACM, IEEE, and CCF.
\end{IEEEbiography}


\begin{IEEEbiography}[{\includegraphics[width=1in,height=1.25in,clip,keepaspectratio]{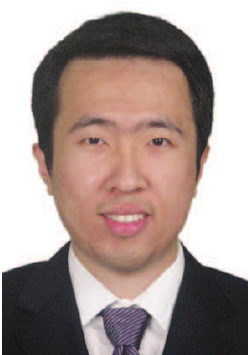}}]{Fan Wu}
is a professor in the Department of Computer Science and Engineering, Shanghai Jiao Tong University. He received his B.S. in Computer Science from Nanjing University in 2004, and Ph.D. in Computer Science and Engineering from the State University of New York at Buffalo in 2009. He has visited the University of Illinois at Urbana-Champaign (UIUC) as a Post Doc Research Associate. His research interests include wireless networking and mobile computing, algorithmic game theory and its applications, and privacy preservation. He has published more than 100 peer-reviewed papers in technical journals and conference proceedings. He is a recipient of the first class prize for Natural Science Award of China Ministry of Education, NSFC Excellent Young Scholars Program, ACM China Rising Star Award, CCF-Tencent ``Rhinoceros bird'' Outstanding Award, CCF-Intel Young Faculty Researcher Program Award, and Pujiang Scholar. He has served as the chair of CCF YOCSEF Shanghai, on the editorial board of Elsevier Computer Communications, and as the member of technical program committees of more than 60 academic conferences. For more information, please visit http://www.cs.sjtu.edu.cn/\texttildelow{}fwu/.
\end{IEEEbiography}

\begin{IEEEbiography}[{\includegraphics[width=1in,height=1.25in,clip,keepaspectratio]{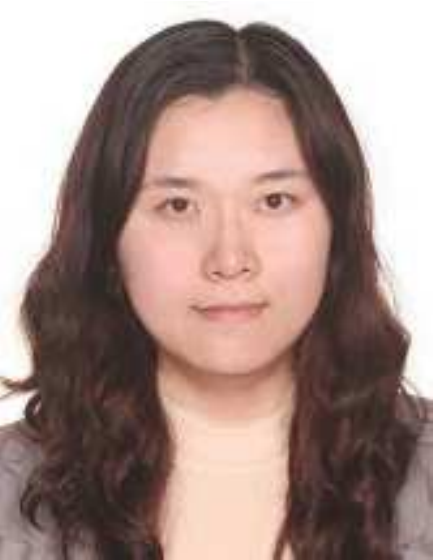}}]{Xiaofeng Gao}
received the B.S. degree in information and computational science from Nankai University, China, in 2004; the M.S. degree in operations research and control theory from Tsinghua University, China, in 2006; and the Ph.D. degree in computer science from The University of Texas at Dallas, USA, in 2010. She is currently an Associate Professor with the Department of Computer Science and Engineering, Shanghai Jiao Tong University, China. Her research  interests include wireless communications, data engineering, and combinatorial optimizations. She has published more than 80  peer-reviewed papers and 6 book chapters in the related area, and she has served as the PCs and peer reviewers for a number of international conferences and journals.
\end{IEEEbiography}

\begin{IEEEbiography}[{\includegraphics[width=1in,height=1.25in,clip,keepaspectratio]{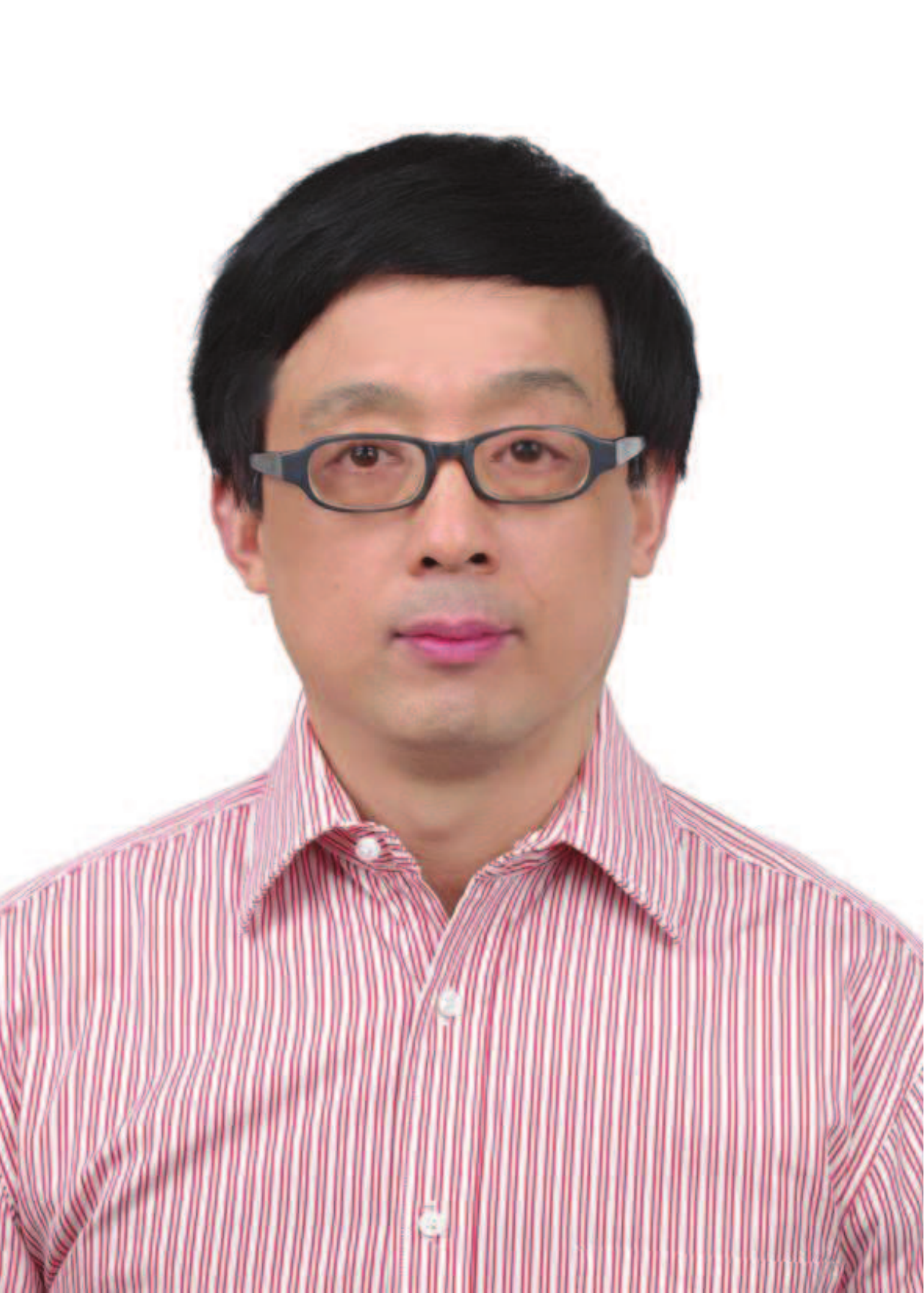}}]{Guihai Chen}
earned his B.S. degree from Nanjing University in 1984, M.E. degree from Southeast University in 1987, and Ph.D. degree from the University of Hong Kong in 1997. He is a distinguished professor of Shanghai Jiaotong University, China. He had been invited as a visiting professor by many universities including Kyushu Institute of Technology, Japan in 1998, University of Queensland, Australia in 2000, and Wayne State University, USA during September 2001 to August 2003. He has a wide range of research interests with focus on sensor network, peer-to-peer computing, high-performance computer architecture and combinatorics. He has published more than 200 peer-reviewed papers, and more than 120 of them are in well-archived international journals such as IEEE Transactions on Parallel and Distributed Systems, Journal of Parallel and Distributed Computing, Wireless Network, The Computer Journal, International Journal of Foundations of Computer Science, and Performance Evaluation, and also in well-known conference proceedings such as HPCA, MOBIHOC, INFOCOM, ICNP, ICPP, IPDPS, and ICDCS.
\end{IEEEbiography}







\end{document}